\documentclass[pre,
 reprint,
 amsmath,amssymb,
 aps,twocolumn, showkeys,
]{revtex4}
\usepackage{wasysym}
\usepackage{color}
\usepackage{graphicx}
\usepackage{dcolumn}
\usepackage{bm}
\usepackage{epsfig,subfigure,float,pseudocode,multirow,footmisc,rotating}
\usepackage[nottoc, notlof, notlot]{tocbibind}
\usepackage{hyperref}
\usepackage{lscape}
\usepackage[a4paper,left=2cm,right=2cm]{geometry}
\usepackage{natbib}
\providecommand{\e}[1]{\ensuremath{\times 10^{#1}}}

\begin{document}

\title{Strains differences in the collective behaviour of zebrafish (\textit{Danio rerio}) in heterogeneous environment}

\author{Axel S\'eguret}
 \email{axel.seguret@univ-paris-diderot.fr}
\affiliation{
 Univ Paris Diderot, Sorbonne Paris Cit\'e\\
 LIED, UMR 8236, 75013, Paris, France
}

\author{Bertrand Collignon}
\affiliation{
 Univ Paris Diderot, Sorbonne Paris Cit\'e\\
 LIED, UMR 8236, 75013, Paris, France
}

\author{Jos\'e Halloy}
 \email{jose.halloy@univ-paris-diderot.fr}
\affiliation{
 Univ Paris Diderot, Sorbonne Paris Cit\'e\\
 LIED, UMR 8236, 75013, Paris, France
}
 
\begin{abstract}
Recent studies show differences in individual motion and shoaling tendency between strains of the same species. Here, we analyse the collective motion and the response to visual stimuli in two morphologically different strains (TL and AB) of zebrafish. For both strains, we observe 10 groups of 5 and 10 zebrafish swimming freely in a large experimental tank with two identical attractive landmarks (cylinders or disks) for one hour. We track the positions of the fish by an automated tracking method and compute several metrics at the group level. First, the probability of presence shows that both strains avoid free space and are more likely to swim in the vicinity of the walls of the tank and the attractive landmarks. Second, the analysis of landmarks occupancy shows that AB zebrafish are more present in their vicinity than TL zebrafish and that both strains regularly transit from one landmark to the other with no preference on the long duration. Finally, TL zebrafish show a higher cohesion than AB zebrafish. Thus, landmarks and duration of the repicates allow to reveal collective behavioural variabilities among different strains of zebrafish. These results provide a new insight into the need to take into account individual variability of zebrafish strains for studying collective behaviour.
\end{abstract}

\keywords{Collective behaviour --- Decision making --- Zebrafish --- Phenotypes}

\maketitle

\section{Introduction}

Collective decision making has been evidenced in many animal species and contexts \cite{Sumpter.2012} including food collection \cite{Detrain.2008}, problem-solving \cite{Dussutour.2009, Jeanson.2010}, collective movement \cite{Petit.2010, Sueur.2011, Radakov.1973, Parrish.2002, Engeszer.2007.1, Herbert-read.2011, Hemelrijk.2012} or nest site selection \cite{Franksetal.2002, Ame.2006}. In this later case, social animals have to select a resting sites among severals potential options in a complex environment. This selection can be made either through individual decisions or complex decision-making processes involving the participation of all individuals \cite{Conradt.Roper.2005} and can be temporary or permanent according to the needs and living style of the considered species.

While this process of collective decision has been studied for a long time in social animals that select a permanent home (social insects for example), only few studies address this problem in the case of nomad species that constantly explore their environment such as birds or fish. In particular, experiments on fish have been generally designed to observe preferences for particular environmental features or landmarks during a relatively short experimental time (few seconds \cite{Ward.2011}, 5 minutes \cite{Sison.2010}, 10 minutes \cite{Miller.2013}, up to 30 minutes per trial \cite{Kistler.2011}). These studies have shown for example that landmarks in a bare tank arouse interest and attract the fish \cite{Kistler.2011, Sullivan.2015} and that variations of the shape of these landmarks can change territory features \cite{Piyumika.2014, Piyumika.2015}.

While these studies provide numerous insights on the individual and collective preferences in fish groups, they generally rely on the comparison between two or more qualitatively different alternatives. Thus, the selection of one option is often based on an intrinsic preference for a particular feature in comparison with the others. Such asymmetric choices may hide the collective decision that results from the internal processes of decision-making of the group.

Here, our aim is to characterise the collective behaviour of groups of zebrafish swimming in an environment with identical landmarks. We observe the collective motion of small shoals of different group sizes (5 and 10 fish) and of two different zebrafish strains (wild type AB or TL). Zebrafish are gregarious vertebrate model organisms to study the cohesion of the group and its decision making \cite{Norton.2010, Oliveira.2013}. Originating from India where they live in small groups in shallow freshwaters \cite{Parichy.2015}, the zebrafish is a diurnal species that prefers staying in groups both in nature and in the laboratory \cite{Engeszer.2007.1, Spence.2008, Mcclure.2006}. To be closer to their natural swimming habits, our experimental method relies on the observation of fish freely swimming in an open environment rather than in a constraining set-up (i.e. maze as used in \cite{Ward.2011, Miller.2013, Sison.2010}). We observe during one hour each group of fish swimming in a large experimental tank with two spots of interest (landmarks). 

The landmarks consist of two striped yellow-green opaque plastic cylinders placed in the water column or two blue transparent floating Perspex disks providing shadow. We choose these colours in the visible spectrum of the zebrafish according to the results of \cite{Risner.2006}. We expect that these landmarks placed in a homogeneous environment could induce a choice of one prefered option by the zebrafish as evidenced for other species faced with identical ressources \cite{Beckers.1990, Beckers.1992, Ame.2006}. On the one hand, we test with cylinders the effect of visual and physical cues in the water column on collective choices. Since zebrafish are known to swim along the walls of experimental tanks, cylinders could act as such walls in the water column. On the other hand, we test with floating disks the impact of visual and physical cues above water, on collective choices. We placed disks and cylinders landmarks to see whether and how the two strains of zebrafish will adapt their group behaviour and their preferences for landmarks.

We quantified behavioural properties of groups of zebrafish from high resolution videos with the help of a tracking system and an automated analysis. Following the method of \cite{Reiser.2009}, we track in real time the positions of the fish and automatically compute their probabilities of presence in the tank, their interindividual distances, the number of individual present near the landmarks as well as the duration of their visits. Then, we compare the distributions obtained for each group size and landmark to highlight differences between strains of zebrafish at the collective level. This methodology based on massive data gathering has now become standard in studies on animal collective behaviour with flies, \textit{Drosophila melanogaster} \cite{Branson.2009, Dankert.2009}, birds, \textit{Sturnus vulgaris} \cite{Ballerini.2007, Miller.2007, Miller.2011} and fish, \textit{Notemigonus crysoleucas} \cite{Strandburg-Peshkin.2013}.


\section{Results}

\subsection{Collective behaviour in heterogeneous environment with cylinders}

\begin{figure}[ht]
\centering
\includegraphics[width=.5\textwidth]{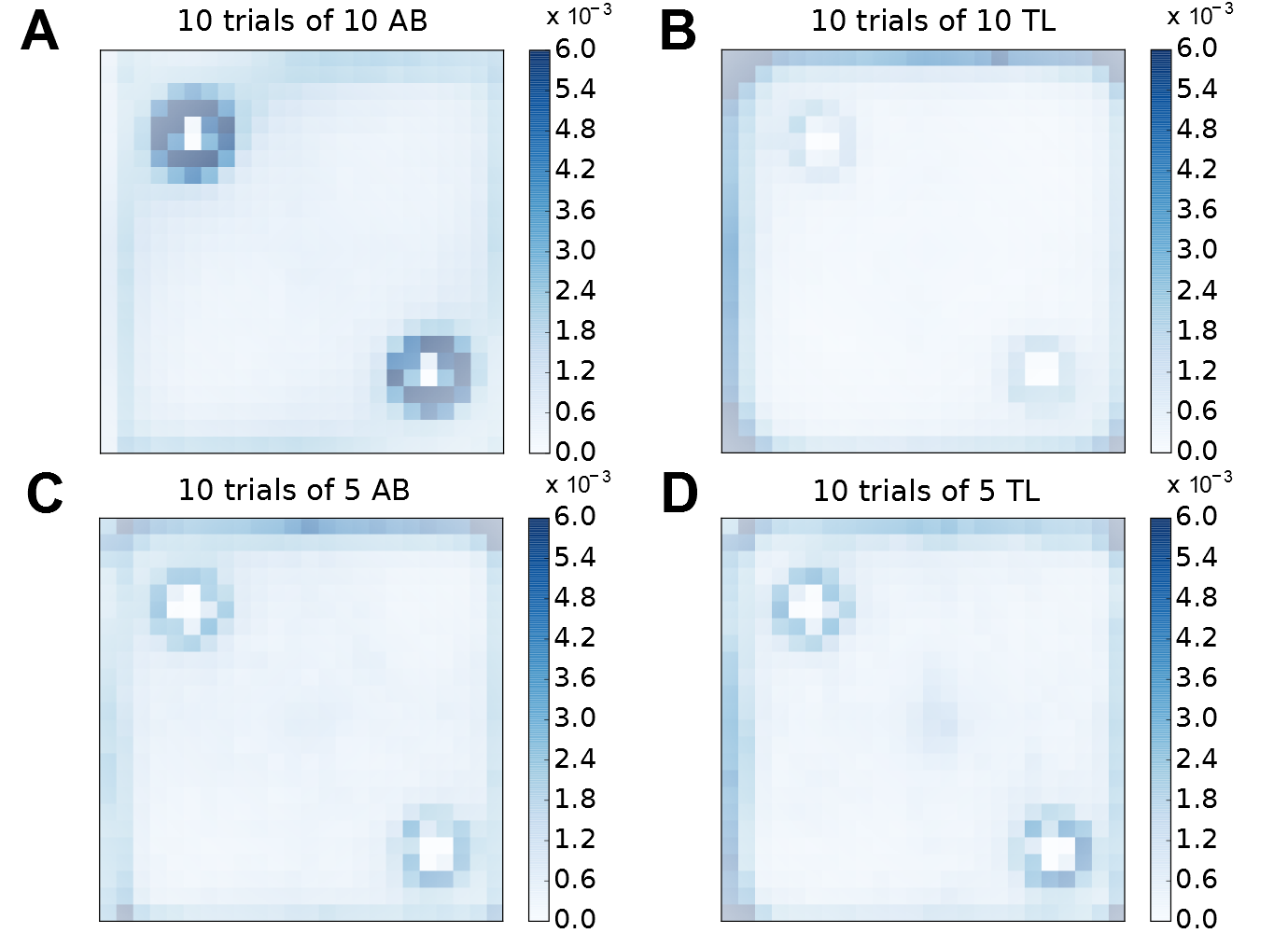}

 \caption{\textbf{Probability of presence} of (A) 10 AB zebrafish, (B) 10 TL zebrafish, (C) 5 AB zebrafish and (D) 5 TL zebrafish in a tank with two cylinders. The probability is calculated on the positions of all zebrafish (i. e. 5 or 10 individuals) observed during one hour and cumulated for 10 trials. The response to the landmarks is strain and group size dependant: while 5 AB and 5 TL zebrafish show similar probability of presence near the cylinders, a larger group size of AB increases the response to the landmarks but decreases the response of groups of 10 TL.}
 \label{fig:all_pdfcylinders}
\end{figure}

In the presence of two striped yellow and green cylinders (see Methods), the groups of 10 AB zebrafish are mainly present along the wall of the tank and around the cylinders, as shown by their average probability of presence computed for the one hour observation period and 10 replicates (Figure~\ref{fig:all_pdfcylinders}). On the contrary, groups of 5 TL, 10 TL and 5 AB zebrafish are less observed near the cylinders but are still present along the walls of the tank (Figure~\ref{fig:all_pdfcylinders}). The probabilities of presence computed for each experiment are shown in the annexe (Figure~\ref{fig:figures1}, Figure~\ref{fig:figures2}, Figure~\ref{fig:figures3} and Figure~\ref{fig:figures4}).

\begin{figure}[ht]
\centering
\includegraphics[width=0.5\textwidth]{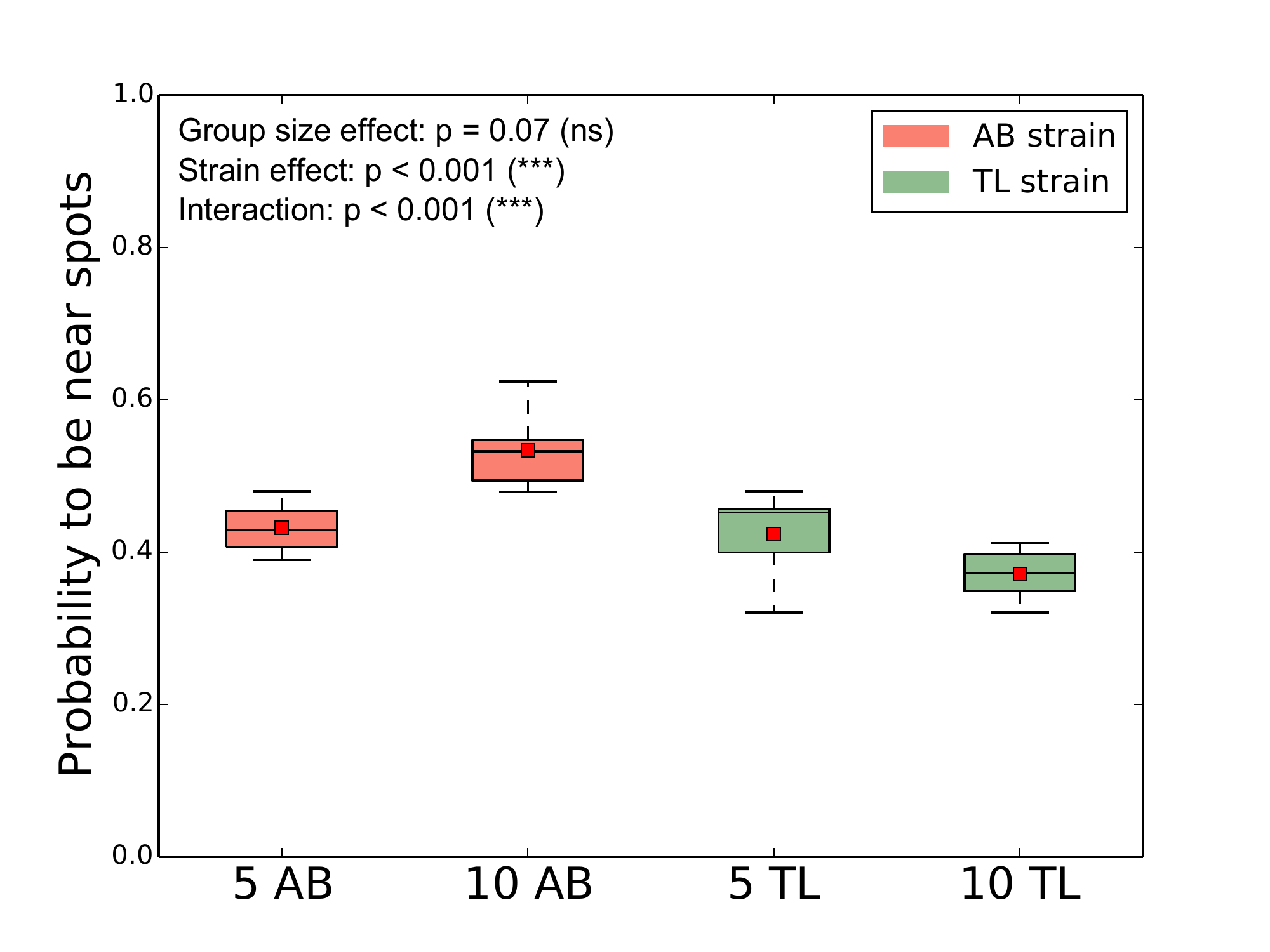}

 \caption{\textbf{Probability to be at 25 cm from the center of the cylinders} for 10 trials with 5 AB, 10 AB, 5 TL or 10 TL zebrafish in a 1m\textsuperscript{2} tank with two cylinders. The black line shows the median, the red square shows the mean. Statistical tests show that groups of 10 AB are more attracted by the cylinders than groups of 5 AB and 5 TL which are more attracted by cylinders than groups of 10 TL. $N$ = 10 measures for 5 AB, 10 AB, 5 TL and 10 TL. * = $p < 0.05$, ** = $p < 0.01$, *** = $p < 0.001$, ns = non significant.}
 \label{fig:anova_cylinders_strains_size}
\end{figure}

To highlight the differential effect of the cylinders on the tested groups, we measured the proportion of positions that were detected at 25 cm from the centre of the cylinders (Figure~\ref{fig:anova_cylinders_strains_size}). This measure confirms that groups of 10 AB zebrafish were more present near the cylinders than groups of 5 AB. In contrast, while groups of 5 TL responded similarly to groups of 5 AB, groups of 10 TL zebrafish were less detected near the cylinders. A two-way ANOVA (group size, strain, $n = 10$ for each experimental conditions) indicated a non-significative effect of the group size ($p = 0.07$, F = 3.59, df = 1) but a significant effect of the strain ($p < 0.001$, F = 43.17, df = 1) and a significant interaction strain/group size ($p < 0.001$, F = 35.38, df = 1) on the attractivity of the cylinders. The interaction effect indicates here a strain-specific effect of the group size on the time spent near the landmarks : groups of 10 AB are more attracted by the cylinders than groups of 5 AB but on the contrary, 10 TL are less detected near the cylinders than groups of 5 TL.

\begin{figure*}[ht]
\centering
\includegraphics[width=0.64\textwidth]{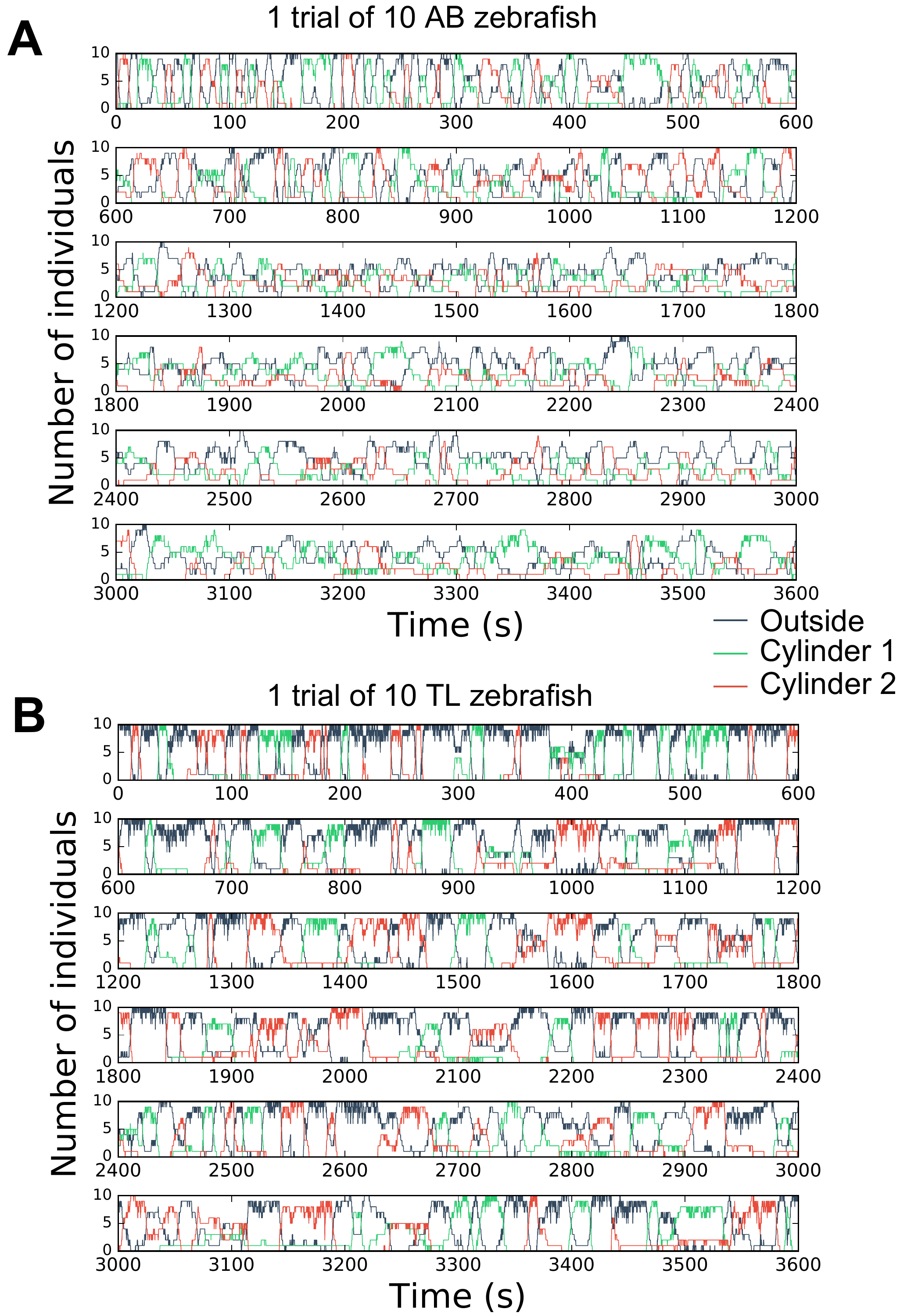}

 \caption{\textbf{Landmark occupancy time series} for 1 trial of 10 individuals of (A) AB zebrafish and (B) TL zebrafish in a tank with two cylinders. For readability, time series are divided in 6 consecutive subplots. Y-axis report the number of individuals. Green line represents individuals in the zone of interest (less than 25 cm away from the center of the landmark) of landmark 1, red line the individuals in the zone of interest of landmark 2 and blue line the individuals outside both zones of interest. The results show that for each strain, the number of fish present in each zone fluctuate in a short time range. The time series of all trials can be seen in the annexe (Figure~\ref{fig:figures7} and Figure~\ref{fig:figures8}).}
 \label{fig:cylindersexploit_temporal}  
\end{figure*}

Then, we studied in more details the dynamics of the presence near the cylinders of the zebrafish swimming in group of 10 individuals. In the AB strain, the fish form a cohesive group that regularly transits from one landmark to the other at the beginning of the trial. Then, the group starts to split in multiple subgroups and the periodicity of the visits becomes less regular (Figure~\ref{fig:cylindersexploit_temporal}A). These oscillations are also observed for groups of the TL strain but contrarily to AB zebrafish, this phenomenon is observed for the whole experimental time (Figure~\ref{fig:cylindersexploit_temporal}B).

\begin{figure*}[ht]
\centering
\includegraphics[width=0.8\textwidth]{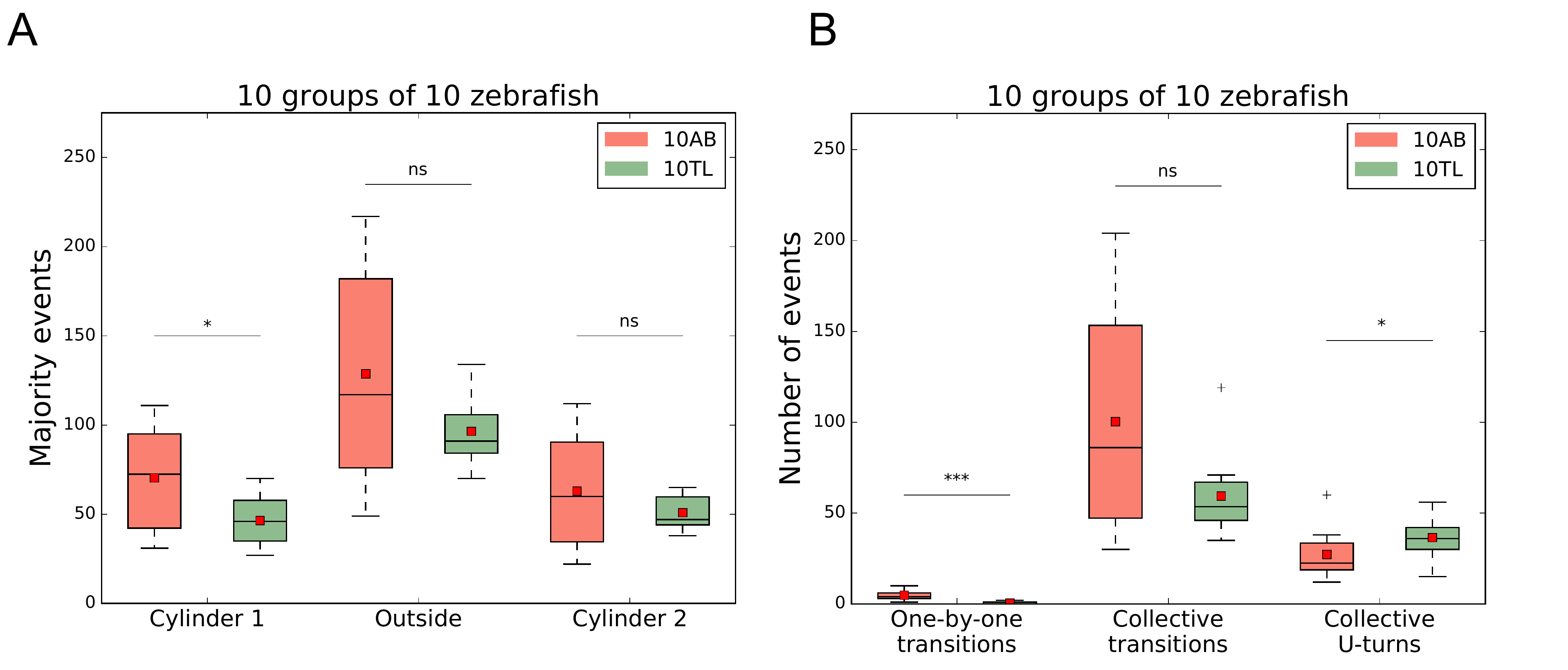}

 \caption{\textbf{Landmark occupancy and transitions} for 10 trials of 10 AB zebrafish and 10 TL zebrafish. (A) Number of majority events occurring around the cylinders and outside. A majority event was considered as soon as more than or equal to 7 fish are aggregated in the same zone. (B) Number of transitions of the majority from one zone to another one. We mainly looked at \textit{"One by one" transitions} (the fish transit one by one from one cylinder to the other), \textit{Collective transitions} (the whole group transits between both cylinders through the outside area) and \textit{Collective U-turns transitions} (the group go back to the previous cylinder). For both strain, while several U-turns were observed, the majority of transitions were made in groups (Collective transitions) and only a few were made one-by-one. The TL strain significantly differs from the AB strain by performing significantly less One-by-one transitions and more Collective U-turns. Each boxplot is composed of 10 values of numbers of events. * = $p < 0.05$, ** = $p < 0.01$, *** = $p < 0.001$, ns = non significant.}
 \label{fig:cylindersexploit_symbdyn}  
\end{figure*}

To quantify the dynamics of these transitions, we computed the number of majority events detected near one of the two landmarks or outside of them. A majority event was counted when 7 or more individuals were simultaneously present in the same zone independently of the duration of this majority event. Figure~\ref{fig:cylindersexploit_symbdyn}A shows that the median and mean number of majority event is always smaller in the TL strain, but this difference is only significant for the majorities detected near one of the cylinder (cylinder 1, Mann-Whitney U test, U = 26, $p = 0.038$ ; cylinder 2, Mann-Whitney U test, U = 48, $p = 0.455$ ; outside, Mann-Whitney U test, U = 38, $p = 0.192$).

Then, we characterised the transitions of the fish from one landmark to the other by analysing the succession of majority events. In particular, we counted the number of "Collective transitions" (two majority events nearby different cylinders separated by a majority outside), the number of "One-by-one transitions" (succession of a majority event in one cylinder and a majority event in the other cylinder) and finally the number of "Collective U-turns" (two majority events in the same cylinder separated by a majority outside). This reveals that the main transitions occurring for AB zebrafish are the collective ones while some collective U-turns and almost no individual transitions were detected (Figure~\ref{fig:cylindersexploit_symbdyn}B red). Similarly, almost no individual transition were observed for the TL zebrafish that perform mainly collective transitions (Figure~\ref{fig:cylindersexploit_symbdyn}B green). TL zebrafish performed also numerous collective U-turns. The absence of individual transitions reveals that both strains are mostly swimming in groups but with different collective dynamics. We compared the results with Mann-Whitney U tests: "One-by-one transitions" (U = 3.0, $p < 0.001$) and "Collective U-turns"  (U = 27.5, $p < 0.050$) between AB and TL are significantly different when "Collective transitions" between AB and TL are not (U = 40, $p = 0.236$). 

\begin{figure*}[ht]
\centering
\includegraphics[width=.6\textwidth]{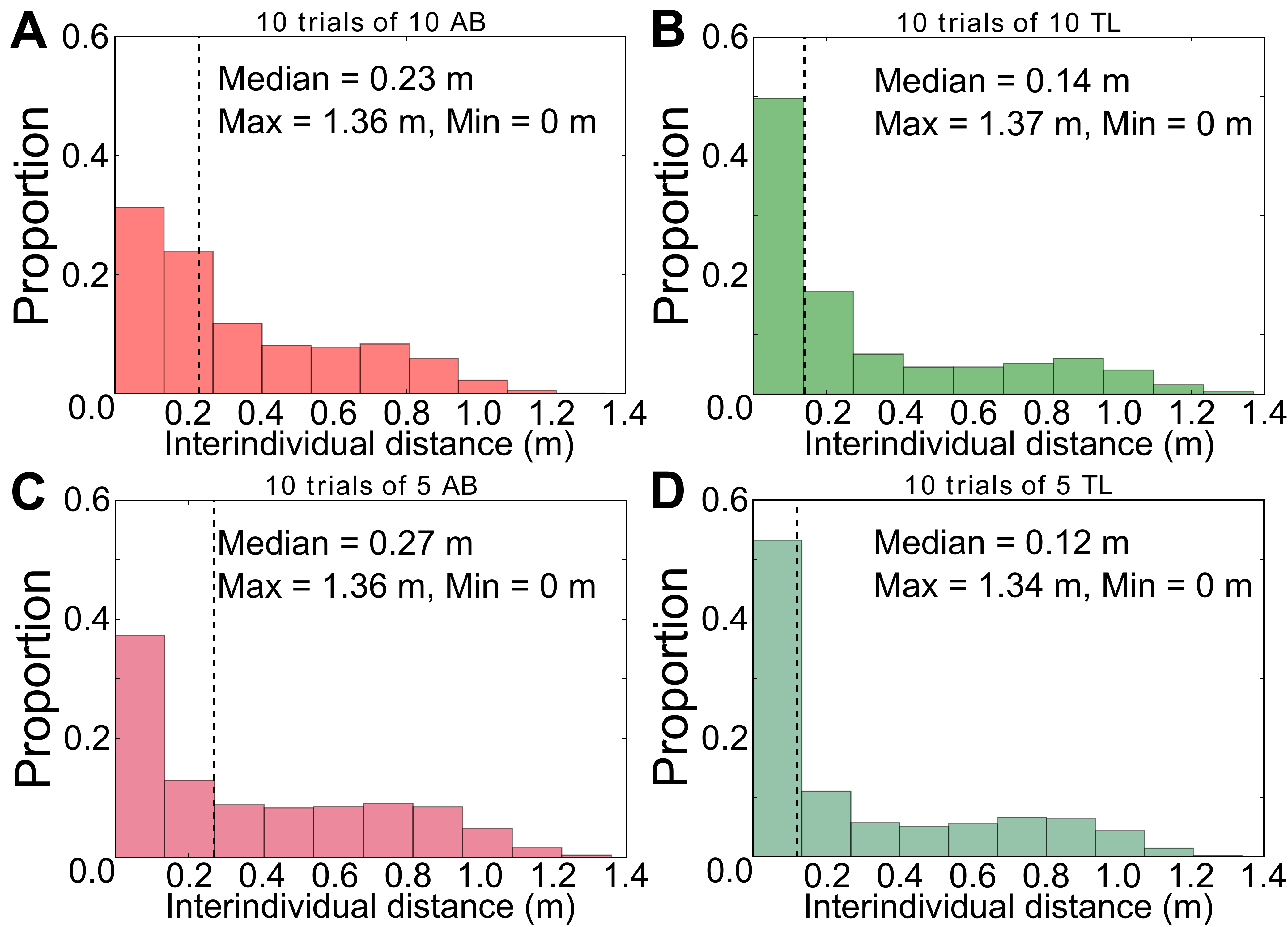}

 \caption{\textbf{Interindividual distances for 10 trials} with (A) 10 AB zebrafish (N = 22 480 363 distances), (B) 10 TL zebrafish (N = 20 981 798 distances), (C) 5 AB zebrafish (N = 5 169 077 distances) and (D) 5 TL zebrafish (N = 4 578 036 distances). N is the number of measured distances. The dashed lines show the medians. The distributions show that groups of TL zebrafish have a stronger cohesion than groups of AB zebrafish. Although group size does not change this distribution in groups of TL strain, it has a strong impact on the AB strain. Distributions of the interindividual distances of 10 AB and 10 TL are different, 5 AB and 5 TL are different, 5 AB and 10 AB are different and 5 TL and 10 TL are different (see in the text the results of the statistical tests).}
 \label{fig:all_interdistdistricylinders}
\end{figure*}

\begin{figure}[ht]
\centering
\includegraphics[width=.5\textwidth]{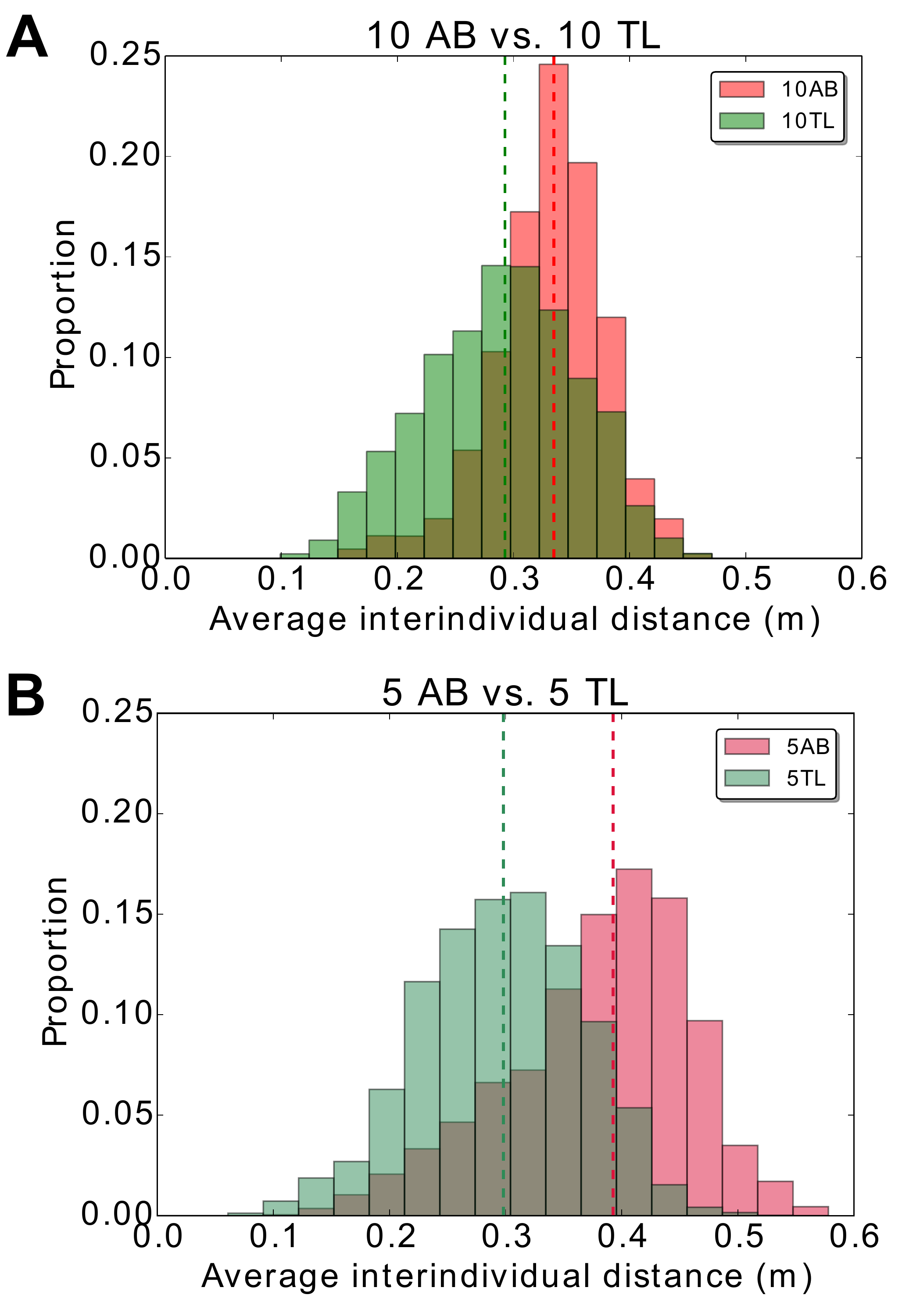}

 \caption{\textbf{Average interindividual distance} (A) for 10 AB zebrafish (N = 518960 measures) versus 10 TL zebrafish (N = 500794 measures). The red distribution represents 10 trials with groups of 10 AB zebrafish and the green represents 10 trials with groups of 10 TL zebrafish ; (B) for 5 AB zebrafish (N = 528357 measures) versus 5 TL zebrafish (N = 495659 measures). The red distribution represents 10 trials with groups of 5 AB zebrafish and the green is for 10 trials with groups of 5 TL zebrafish. Dashed lines represent medians. TL zebrafish are more cohesive than AB zebrafish. Smaller groups of AB zebrafish show a shift to higher values. The distributions of the average interindividual distances of 10 AB and 10 TL are different, 5 AB and 5 TL are different, 5 AB and 10 AB are different and 5 TL and 10 TL are different.}
 \label{fig:all_averageinterdistdistibutioncylinders}
\end{figure}

Finally, we computed the interindividual distances between all fish to characterize the cohesion of the group for both strains and group sizes. The distribution of all interindividual distances (Figure~\ref{fig:all_interdistdistricylinders}) shows that groups of TL zebrafish have a stronger cohesion (5 TL: Median = 0.12m and 10 TL: Median = 0.14m) than the groups of AB zebrafish (5 AB: Median = 0.27m and 10 AB: Median = 0.23m). The intra-strain comparison for the two group sizes shows that the distribution significantly differs from each other (Kolmogorov-Smirnov test, 5 AB vs 10 AB, D = 0.102, $p < 0.001$; 5 TL vs 10 TL, D = 0.080, $p < 0.001$). The inter-strain comparison for similar group sizes also reveals a statistical difference between the distributions (Kolmogorov-Smirnov test, 10 AB vs 10 TL, D = 0.184, $p < 0.001$; 5 AB vs 5 TL, D = 0.161, $p < 0.001$). The distributions of the average interindividual distance measured at each time step confirms these results (Figure~\ref{fig:all_averageinterdistdistibutioncylinders},  Kolmogorov-Smirnov test, 5 AB vs 10 AB, D = 0.433, $p < 0.001$; 5 TL vs 10 TL, D = 0.051, $p < 0.001$; 10 AB vs 10 TL, D = 0.333, $p < 0.001$; 5 AB vs 5 TL, D = 0.464, $p < 0.001$. In addition, the analysis of the evolution of the average interindividual distance revealed that the cohesion of the fish decreases for both strains and both group sizes through the time (Figure~\ref{fig:figures11} of the annexe).

\clearpage
\subsection{Collective behaviour in heterogeneous environment with disks}

\begin{figure}[ht]
\centering
\includegraphics[width=.5\textwidth]{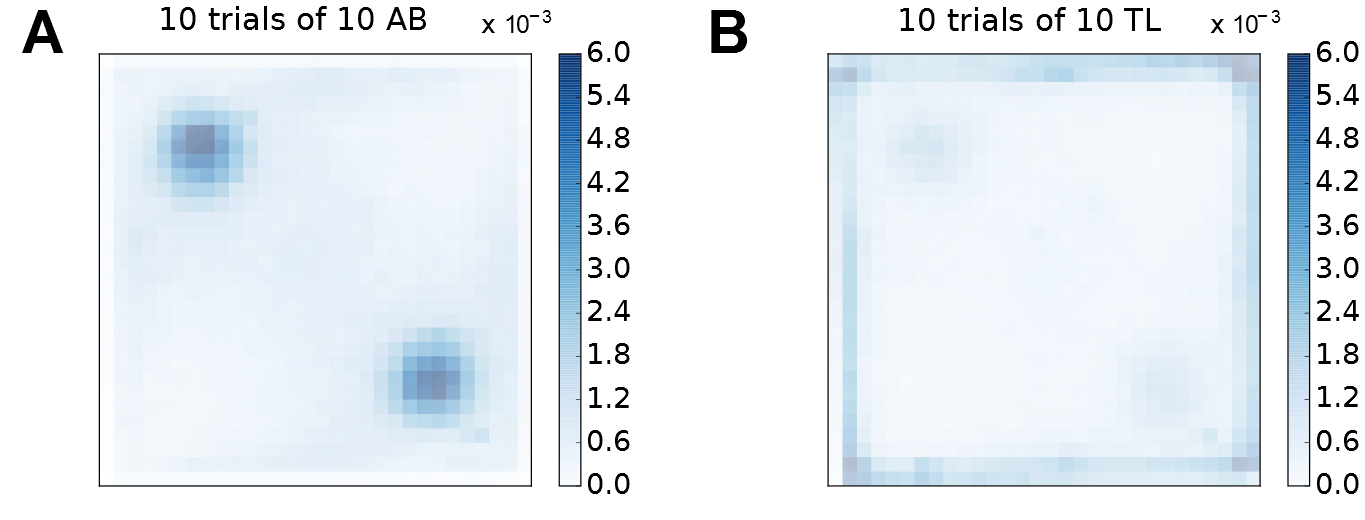}

 \caption{\textbf{Probability of presence} of (A) 10 AB zebrafish in a tank with two disks, (B) 10 TL zebrafish in a tank with two disks. Attractivity to landmarks is strain dependant: the probability of presence of finding 10 AB zebrafish around the landmarks is 2 times higher than that of 10 TL zebrafish.}
 \label{fig:all_pdfdisks}  
\end{figure}

\begin{figure}[ht]
\centering
\includegraphics[width=0.5\textwidth]{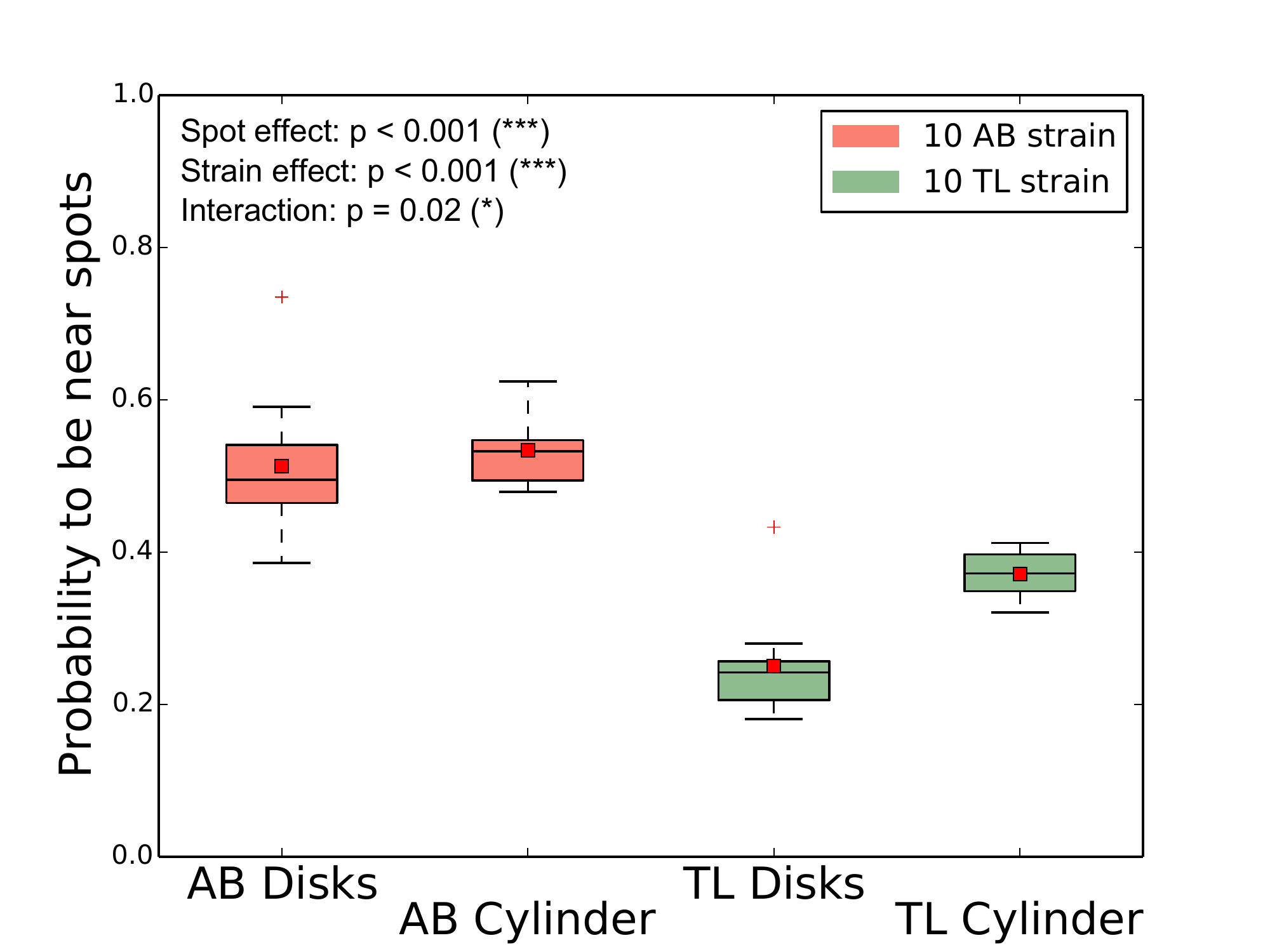}

 \caption{\textbf{Probability to be at 25 cm from the center of cylinders or disks} for 10 trials. Groups of 10 AB or 10 TL zebrafish in a tank with two cylinders or two disks. The black line markes the median, the red square markes the mean. A serie of tests shows that groups of 10 AB are attracted by cylinders as much as disks, groups of 10 TL are more attracted by cylinders than disks, groups of 10 AB are more attracted by the cylinders and the disks than groups of 10 TL. N = 10 measures for AB disk, AB cylinder, TL disk and TL cylinder. * = $p < 0.05$, ** = $p < 0.01$, *** = $p < 0.001$, ns = non significant.}
 \label{fig:anova_cylinders_disks_strains_size}
\end{figure}

\begin{figure}[ht]
\centering
\includegraphics[width=.35\textwidth]{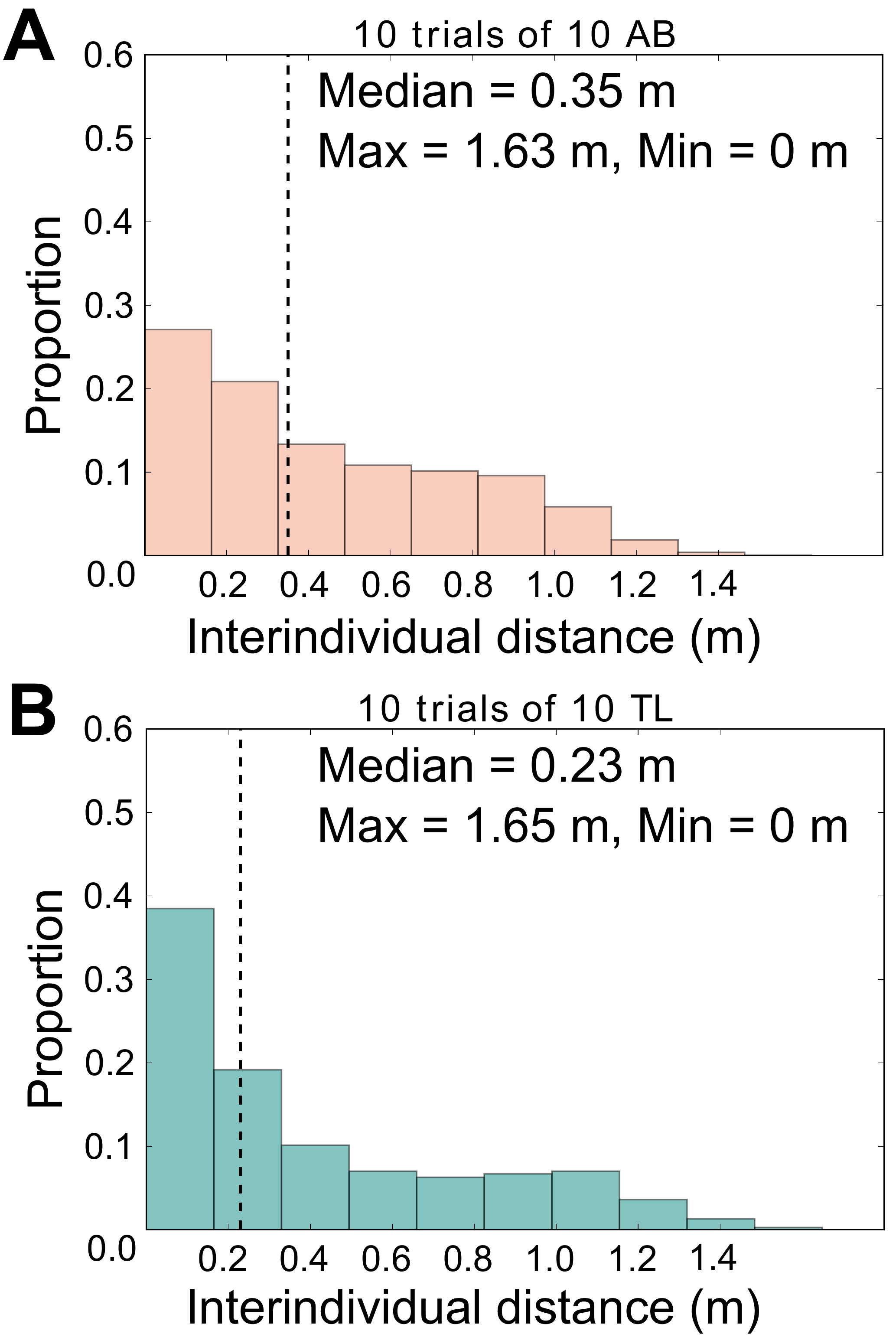}

 \caption{\textbf{Interindividual distances for 10 trials} with groups of (A) 10 AB zebrafish (N = 1 620 450 distances), (B) 10 TL zebrafish (N = 1 620 450 distances) in a tank with two disks. N is the number of measure of distance. Distribution of the interindividual distance shows that groups of TL zebrafish have a stronger cohesion than groups of AB zebrafish. Max and min show the maximum or minimum distances found between fish. A Kolmogorov-Smirnov test shows that the distributions of the interindividual distances of 10 AB and 10 TL are different.}
 \label{fig:all_interdistdistridisks}
\end{figure}

We also placed symmetrically two floating identical perspex transparent blue disks, at 25$\sqrt{2}$ cm from two opposite corners along the diagonal. Acting as roofs on the water, they create shades that could attract zebrafish. We did 10 trials on groups of 10 AB and 10 TL zebrafish. We observed similar behaviours than in the previous experiments with cylinders. The maximum probability of presence under disks with AB zebrafish reaches 4.5\e{-3} when for TL zebrafish it reaches only 1\e{-3} (Figure~\ref{fig:all_pdfdisks}). Again, TL zebrafish spent the majority of their time near the borders of the tank (the probabilities of presence for each experiment are shown in the Figure~\ref{fig:figures5} and Figure~\ref{fig:figures6} of the annexe). We compared the probability to be near the spots for both strains and both types of landmarks with a two-way ANOVA (Figure~\ref{fig:anova_cylinders_disks_strains_size}). It revealed that the type of landmarks  ($p < 0.001$, F = 11.37, df = 1) and the strain of zebrafish affect the attraction ($p < 0.001$, F = 102.95, df = 1) while there is also an evidence of an interaction effect between type of landmarks and strains ($ p = 0.02$, F = 5.67, df = 1). Thus, while groups of 10 AB are attracted by cylinders as much as by disks, groups of 10 TL are more attracted by cylinders than disks. Eventually, TL zebrafish were also significantly more cohesive than the AB zebrafish in the presence of the floating disks, as shown by the distribution of the interindividual distances (medians for 10 AB : 0.35m, for 10 TL : 0.23m, Figure~\ref{fig:all_interdistdistridisks}, Kolmogorov-Smirnov test, D = 0.135 and $p < 0.001$).

\section{Discussion}

We investigated whether collective motion and collective choice can differ in groups of 5 and 10 individuals of the same species (zebrafish \textit{Danio rerio}) but of different strains (AB versus TL). Both strains were laboratory wild types, had the same age and were raised under the same conditions. We observed zebrafish swimming for one hour in the presence of two identical landmarks (immersed cylinders or floating disks) that were put symmetrically in the tank. 
Although one hour observations show that the zebrafish do not select one of the two landmarks, the fish were mainly swimming together and oscillate from one landmark to the other with short resting times. Thus, while all individuals can be punctually present at the same landmark (Figure~\ref{fig:cylindersexploit_temporal}, Figure~\ref{fig:figures7} and Figure~\ref{fig:figures8} of the annexe), the probability of presence computed for the entire experimental time shows that the fish were equally present at both stimuli. Therefore, no collective choice emerged on the long time for both strains of zebrafish and group size. Hence, long time and short time range analyses reveal opposite results and conclusions on collective motions.

Our methodology is complementary to typical Y-maze experiments. We extend and compare their conclusions to our observations with repeated interactions between the fish and their environment. During a hour, the collective behaviour of zebrafish contrasts with other collective species in which spatial fidelity emerges from the interactions between the individuals that take place in the resting sites (in cockroaches \cite{Ameetal.2004}, in hymenoptera \cite{Franksetal.2002}). These oscillations from one site to the other could originate from individual differences among group members: \textit{bold} and \textit{shy} behavioural profiles have been evidenced in zebrafish according to the intrinsic propensity of each fish to explore new environments \cite{Dahlbometal.2011}. It also has been identified in other fish species \cite{Harcourt.2009}. In this context, the presence of bolder fish in the group could favour the transition from one spot to the other while groups composed only by shy individuals could show less frequent departures \cite{LeblondAndReebs2006}.

A more detailed analysis show quantitative differences among the two studied strains and group sizes. Concerning the response of the fish to the landmarks, we highlight that groups of 5 AB and TL zebrafish show the same attraction for the cylinders by computing the probability for the fish to be observed near these landmarks. This attraction increases for groups of 10 AB zebrafish but decreases for groups of 10 TL zebrafish. This strain difference is also observed in the experiments with floating disks. In addition, the type of the landmarks seems to be determinant for TL zebrafish as they prefer objects immersed in the water column than objects lying on the surface of the water. Hence, it exists a difference of collective behaviour between the two tested strains of zebrafish.

This different response to heterogeneities may be based on the intrinsic preference of the fish of a particular strain for congeners or for landmarks. Such difference has already been shown in shoaling tendency between several strains of guppies \cite{Magurran.1995} and zebrafish \cite{Wright.2003}, \cite{Wright.2006}. In their natural environment, fish have to balance the costs of risks and benefits of moving in groups or staying near landmarks \cite{Millot.2009}. Moving in groups first of all prevents the fish to be static preys, and second allow spatial recognition and easier food and predators detection. The drawback is that the takeover of the fish on the territory is punctual and they have less chance to find areas where they can hide. Staying around landmarks gives the fish a feeling of control of the territory and the possibility to hide from predators. In that case the drawback is that the preys will rarely cross the territory of the fish.

Regarding the structure of the group, we notice that whatever the group size of TL zebrafish, the cohesion of the group does not change and is always stronger than those of groups of AB zebrafish. Also, the bigger the group of AB zebrafish, the stronger the cohesion. These differences of group cohesion may be based on physical features differences between AB and TL zebrafish. Cohesion differences could be explained by phenotype differences between the two strains. Some studies demonstrated that a large variability exists in the individual motion and shoaling tendency of the zebrafish according to their age or strain. For example, adults AB and casper zebrafish swim longer distance than ABstrg, EK, TU or WIK zebrafish \cite{Lange.2013}. Likewise, the interindividual distance between shoal members decreases from 16 body length to 3.5 body length between day 7 and 5 months after fertilization \cite{Buske.2011}. Also, it has been demonstrated that the fin size has an impact on the swimming performance and behaviour of the zebrafish \cite{Plaut.2000}. Starting from the assumptions that AB and TL zebrafish have the same visual acuity and the same lateral line performance, the only source of separation between both strains are their fin lengths and the patterns on the skin: TL zebrafish are homozygous for leo$^{t1}$ and lof$^{dt2}$, where leo$^{t1}$ is a recessive mutation causing spotting in adult zebrafish and lof$^{dt2}$ is a dominant homozygous viable mutation causing long fins \cite{Iovine.2000}, \cite{Watanabe.2006}. Thus, AB zebrafish have short fins and TL zebrafish show long fins. It may suggest that TL zebrafish move a higher quantity of water with their long fins when swimming and thus emit a stronger signal of presence (hydrodynamical signal). Thus, it may be easier for conspecifics in the moving shoal to perceive the signal through their lateral line and realign themselves according to their conspecifics. If the realignment becomes easier, it is simpler for TL zebrafish to keep their position in the shoal, which increases its cohesion. Following a similar hypothesis, the signal of presence is weaker for AB zebrafish due to their shorter fins. Thus, realignment in the moving shoal is less performant and their cohesion decreases. Each of the two hypotheses could explain the collective behaviours observed during the experiments and nothing prevents merging both of them.

In conclusion, this study demonstrates that behavioural differences exist at the individual and collective levels in the same species of animal. The analysis of the dynamics reveals that AB and TL zebrafish mainly oscillate in groups between landmarks. In addition, increasing the size of the group leads to opposite results for the two strains: groups of 10 AB zebrafish are proportionally more detected near the landmarks than groups of 5 AB while groups of 10 TL zebrafish are less attracted by the landmarks than groups of 5 TL. Finally, the two tested zebrafish strains show differences at the structural level: (1) groups of TL zebrafish are more cohesive than groups of AB zebrafish and (2) AB zebrafish collective responses to landmarks show that they are generally more present near the cylinders and floating disks than TL zebrafish. Thus, this study provides evidences that zebrafish do not select resting site on the midterm and highlights behavioural differences at the individual and collective levels among the two tested strains of zebrafish. Future studies of collective behaviour should consider the tested strains, the intra-strain composition of the shoals and the duration of each trial.


\section{Methods}

\subsubsection{Ethics statement}

The experiments performed in this study were conducted under the authorisation of the Buffon Ethical Committee (registered to the French National Ethical Committee for Animal Experiments \#40) after submission to the state ethical board for animal experiments.

\subsubsection{Fish and housing}

We acquired 500 adult wild-type zebrafish (200 AB strain and 300 TL strain) from Institut Curie (Paris) and raised them in tanks of 60L by groups of 50. The zebrafish AB line show a zebra skin, short tail and fin. The zebrafish TL line show a spotted skin, long tail and fin and barbel. Both strains are 3.5 cm long. The zebrafish used for the experiments are adult fish between 5 months and 18 months of age. During this period, zebrafish show a shoaling tendency allowing study of their collective behaviours. We kept fish under laboratory condition, $27\,^{\circ}{\rm C}$, 500$\mu$S salinity with a 9:15 day:night light cycle. The fish were reared in 55 litres tanks and were fed two times per day (Special Diets Services SDS-400 Scientic Fish Food). Water pH is maintained at 7.5 and nitrites (NO$^{2-}$) are below 0.3 mg/l. We measured the size of the caudal fins of 10 AB (about 0.4 cm) and 10 TL (about 1.1 cm) zebrafish.

\begin{figure*}[ht]
\centering
\includegraphics[width=0.9\textwidth]{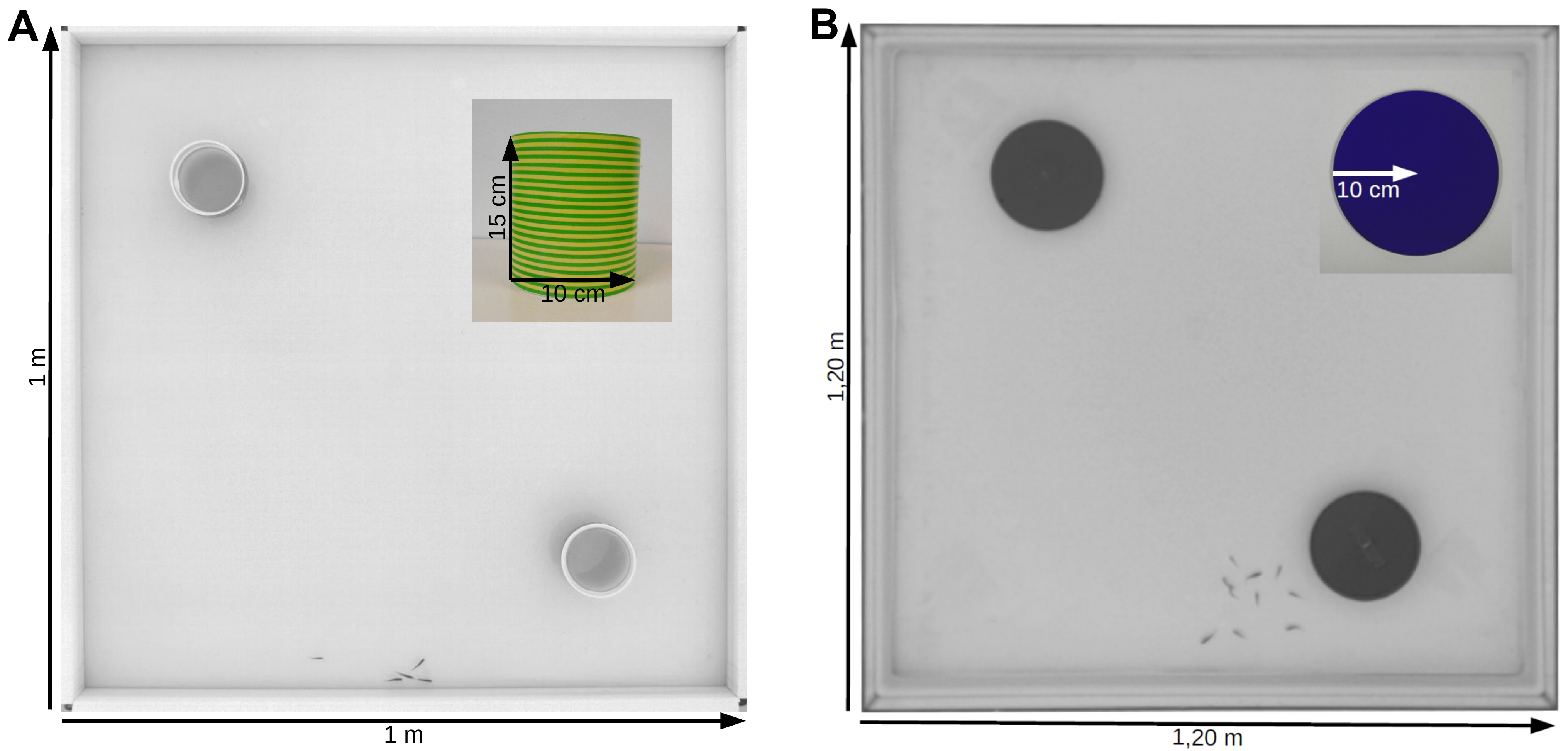}
\caption{(A) Experimental setup (1m x 1m) with two cylinders symmetrically placed and a magnified cylinder ($\phi$ = 10 cm, Height = 15 cm). (B) Experimental setup (1.2m x 1.2m) with two blue perspex disks symmetrically placed and a disk ($\phi$ = 20 cm). In the experimental tank, the water column is 6 cm. Luminosity is ensured by 4 fluorescents lamps of 80W placed on each side of the tank, on the floor and directed towards the walls to provide indirect lightning. The whole set-up is confined in a 2 m x 2 m x 2.35 m experimental chamber (cage) surrounded by white sheets to isolate the experiments and to homogenise luminosity.}
\label{fig:set-up}
\end{figure*}

\subsubsection{Experimental setup}

The experimental tank consists in a 1.2 m x 1.2 m tank confined in a 2 m x 2 m x 2.35 m experimental area surrounded by a white sheet, in order to isolate the experiments and homogenise luminosity. The wall of the experimental tank were covered with white tape and the water column is 6 cm. Water pH is maintained at 7.5 and Nitrites (NO$^{2-}$) are below 0.3 mg/l. The experiments with disks were performed in the experimental tank while those with cylinders were performed in a white square arena (1 m x 1 m x 0.15 m) placed in the experimental tank. Groups of zebrafish were randomly formed at the beginning of the experiments.

The experiments were recorded by a high resolution camera (2048 x 2048 px, Basler Scout acA2040-25gm) placed above the experimental tank and recording at 15 fps (frame per second). Luminosity is ensured by 4 fluorescents lamps of 80W placed on each side of the tank, on the floor and directed towards the walls to provide indirect lightning.

To trigger interest of fish, we placed symmetrically in the set-up either two floating disks ($\phi$ = 20 cm) or two cylinders ($\phi$ = 10 cm, Height = 15 cm) surrounded by yellow and green striped tape \cite{Gerlai.2014.1, Lau.2011, Saverino.2008}. To avoid the presence of a blind zone, the cylinders were slightly tilted toward the centre of the tank. The center of both disks and cylinders are at 25$\sqrt{2}$ cm from two opposite corners along the diagonal of the tank (Figure~\ref{fig:set-up}).

\subsubsection{Experimental procedure}

We recorded the behaviour of zebrafish swimming in the experimental tank during one hour. Before the trials, the attractive landmarks are put in the setup and fish are placed with a hand net in a cylindrical arena (20 cm diameter) made of Plexiglas placed in the centre of our tank. Following a five minutes acclimatisation period, this arena is removed and the fish are able to freely swim in the experimental arena. We performed 10 trials for each strain with the floating disks and 10 trials for each combination of parameters (number of fish x strain) with the cylinders for a total of 60 experiments. Each fish was never tested twice in the same experimental condition.

\subsubsection{Tracking and Data analysis}

The experiments with cylinders were recorded at 15 fps and tracked online by a custom made tracking system based on blob detection. We call a batch a group of 10 experiments. For these batchs, each experiment consists of 540000 positions (10 zebrafish x 54000 frames) and 270000 positions (5 zebrafish x 54000 frames). For experiments with disks, we faced tracking troubles. Since the fish below the floating disks were difficult to distinguish by the program due to a lack of sufficient contrast, experiments with floating disk were tracked offline by two custom Matlab scripts. A first script automatically identifies the positions of the fish swimming outside of the floating disks by blob detection. Since this method did not allow a perfect detection of all the individuals, we developed a second script that was run after the first one and that plotted the frame where a fish (or more) was undetected for the user to manually identify the missing individual(s). It allowed us to identify the fish that were partially hidden during a collision/superposition with another fish or the fish that were situated under the floating disks. Since this analysis tool is time-costly, we only analysed 1 fps for all experiments with disks. For these batchs, each experiment consists of 36000 positions (10 zebrafish x 3600 positions).

Since our tracking system did not solve collision with accuracy, we did not calculate individual measures but characterised the aggregation level of the group. The probability of presence of the fish was calculated by the cumulated positions of all individuals along the entire experiment. We also calculated the distance between the fish and the attractive landmarks as well as the inter-individual distances between the fish and the average inter-individual distance. Finally, we computed the time of shelter occupancy as the time that is spent by the fish at less than 25cm of the attractive landmarks. These time sequences were calculated according to the number of fish present near the landmarks. All scripts were coded in Python using scientific and statistic libraries (numpy, pylab, scilab and matplotlib).

To compute the number of majority events, the number of fish was average over the 15 frames of every second. This operation garanties that a majority event is ended by the departure of a fish and not by an error of detection during one frame by the tracking system. Figure~\ref{fig:figures9} and Figure~\ref{fig:figures10} of the annexe show the proportions of the durations of the the majority events before and after this interpolation. 

\begin{table}
\begin{tabular}{|c|c|}
  \hline
Experiments & Mean of percentages of tracking efficiencies\\
  \hline
10 AB cylinders & 96.05\%\\
  \hline
10 TL cylinders & 92.73\%\\
  \hline
5 AB cylinders & 97.80\%\\
  \hline
5 TL cylinders & 91.77\%\\
  \hline
10 AB disks & 100\%\\
  \hline
10 TL disks & 100\%\\
  \hline
\end{tabular}
\caption{\textbf{Means of percentages of tracking efficiencies}.}
\label{fig:efficiency}
\end{table}

\subsubsection{Statistics}

For the Figures~\ref{fig:anova_cylinders_strains_size} and ~\ref{fig:anova_cylinders_disks_strains_size}, 10 measures of means of the probability for different groups of zebrafish to be near the landmarks are ploted. They have been tested using a two-way ANOVA. We then compared the data between each group using a one-way ANOVA and finally used a Tukey's honest significant difference criterion post-hoc test. We did these tests on MATLAB and chose 0.001 as significance level.
In the Table~\ref{fig:efficiency}, we show the number of majority events after interpolation of the data at 1 fps. This table is related to the Figure~\ref{fig:cylindersexploit_symbdyn}. We used Mann-Whitney U tests to compare the number of events between strains, areas and transition types. These tests are performed on 10 values of majority events for each strain, area and transition type. These tests were made with the python package scipy. We chose 0.001 (***), 0.01 (**) and 0.05(*) as significance levels.
For the Figure~\ref{fig:all_interdistdistricylinders} as well as the Figures~\ref{fig:all_averageinterdistdistibutioncylinders} and~\ref{fig:all_interdistdistridisks}, we compared the distribution with Kolmogorov-Smirnov tests. These tests were made with the python package scipy. We chose 0.001 as significance level.

\phantomsection
\section*{Acknowledgments}

The authors thank Filippo Del Bene (Institut Curie, Paris, France) who provided us the fish observed in the experiments reported in this paper. This work was supported by European Union Information and Communication Technologies project ASSISIbf, FP7-ICT-FET-601074. The funders had no role in study design, data collection and analysis, decision to publish, or preparation of the manuscript.

\addcontentsline{toc}{section}{Acknowledgments}

\clearpage
\onecolumngrid
\section{Annexe}

Supplementary figures of "Strains differences in the collective behaviour of zebrafish (\textit{Danio rerio}) in heterogeneous environment".

\begin{figure*}[ht]
\centering
\includegraphics[width=0.9\textwidth]{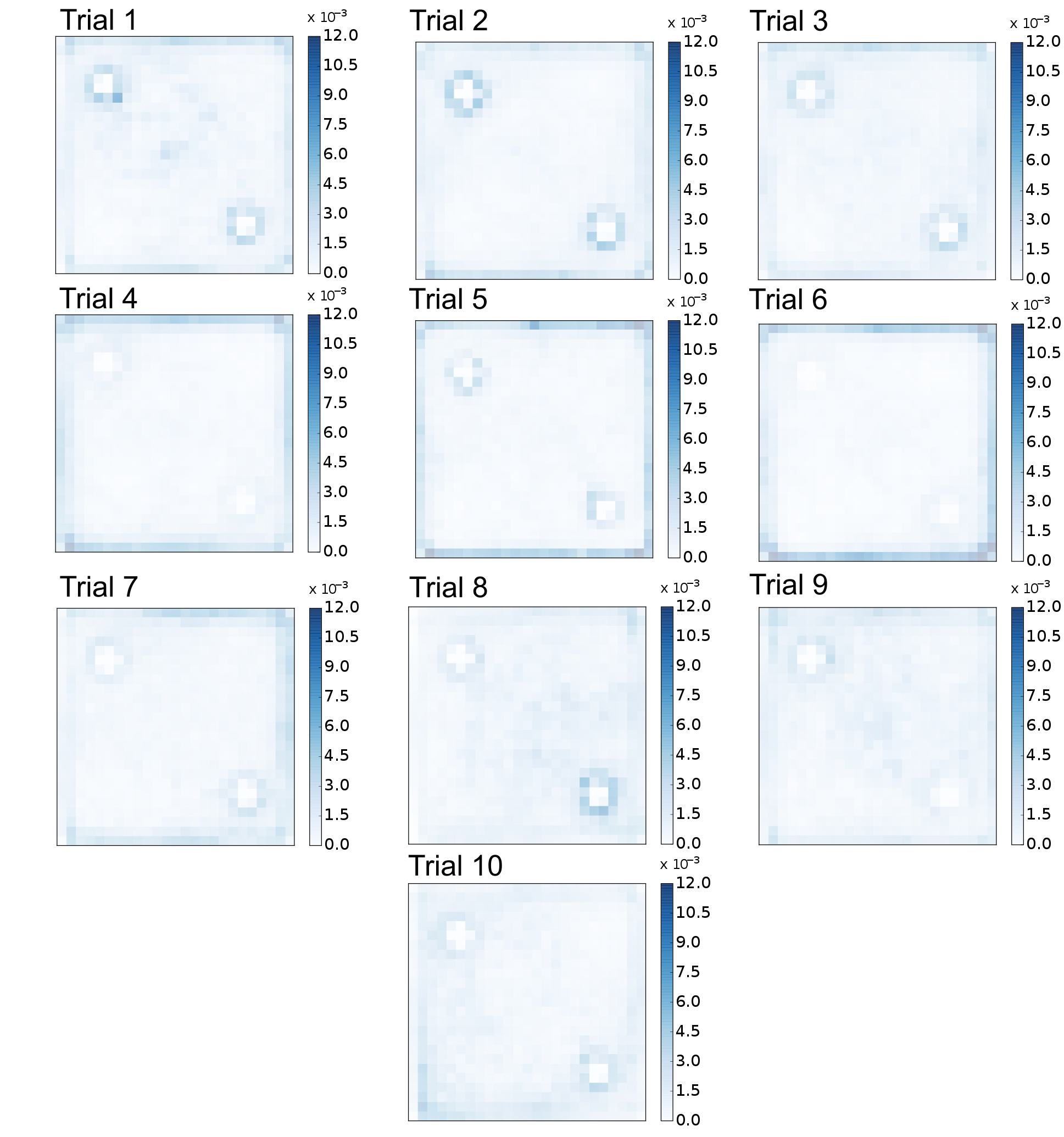}
\caption{\textbf{Probabilities of presence} of 10 trials of 5 AB zebrafish with two cylinders. The probability is calculated on the positions of all zebrafish observed during one hour. The bluer the more fish detected.}
\label{fig:figures1}
\end{figure*}

\begin{figure}[ht]
\centering
\includegraphics[width=1\textwidth]{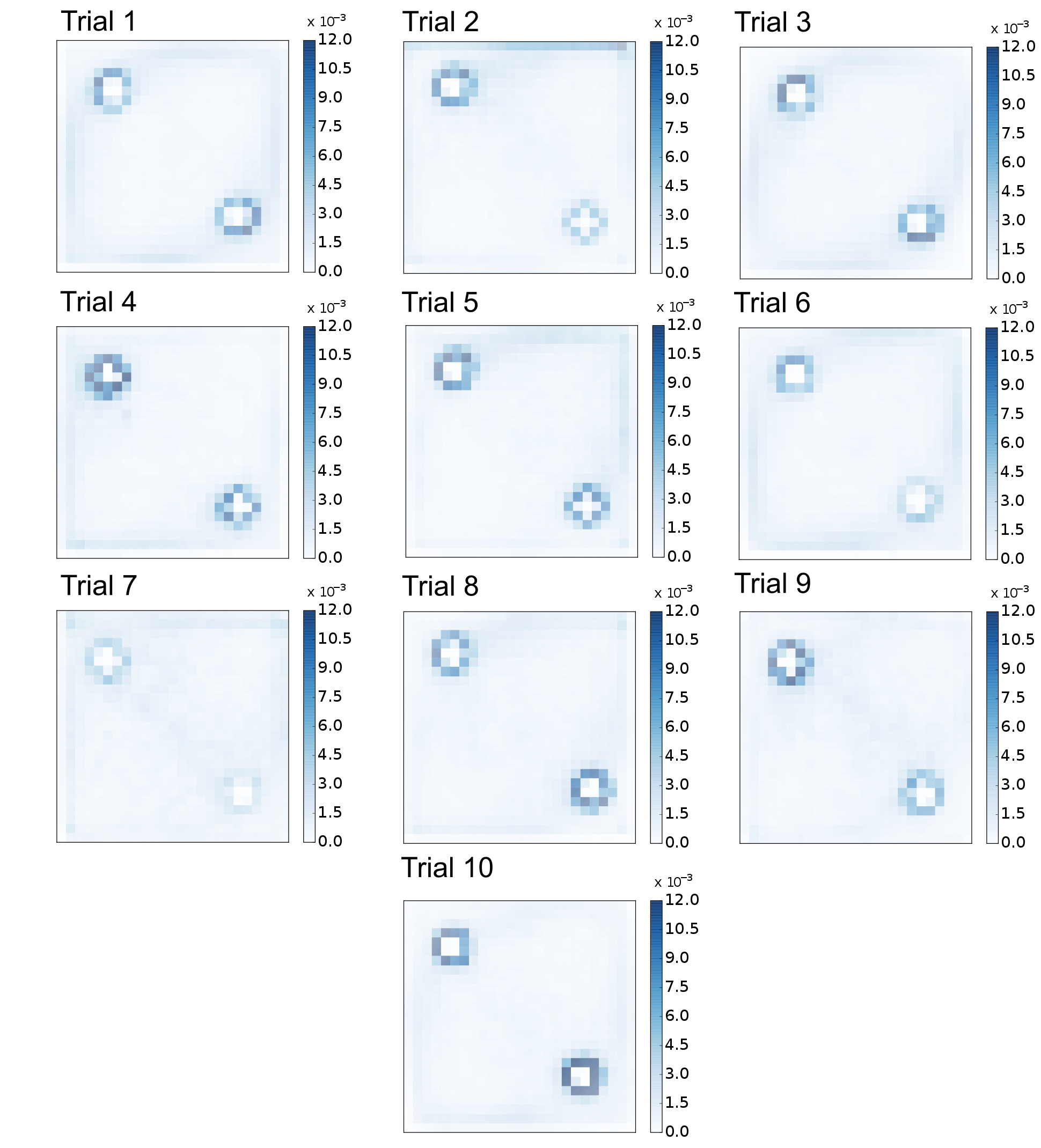}
\caption{\textbf{Probabilities of presence} of 10 trials of 10 AB zebrafish with two cylinders. The probability is calculated on the positions of all zebrafish observed during one hour. The bluer the more fish detected.}
\label{fig:figures2}
\end{figure}

\begin{figure}[ht]
\centering
\includegraphics[width=1\textwidth]{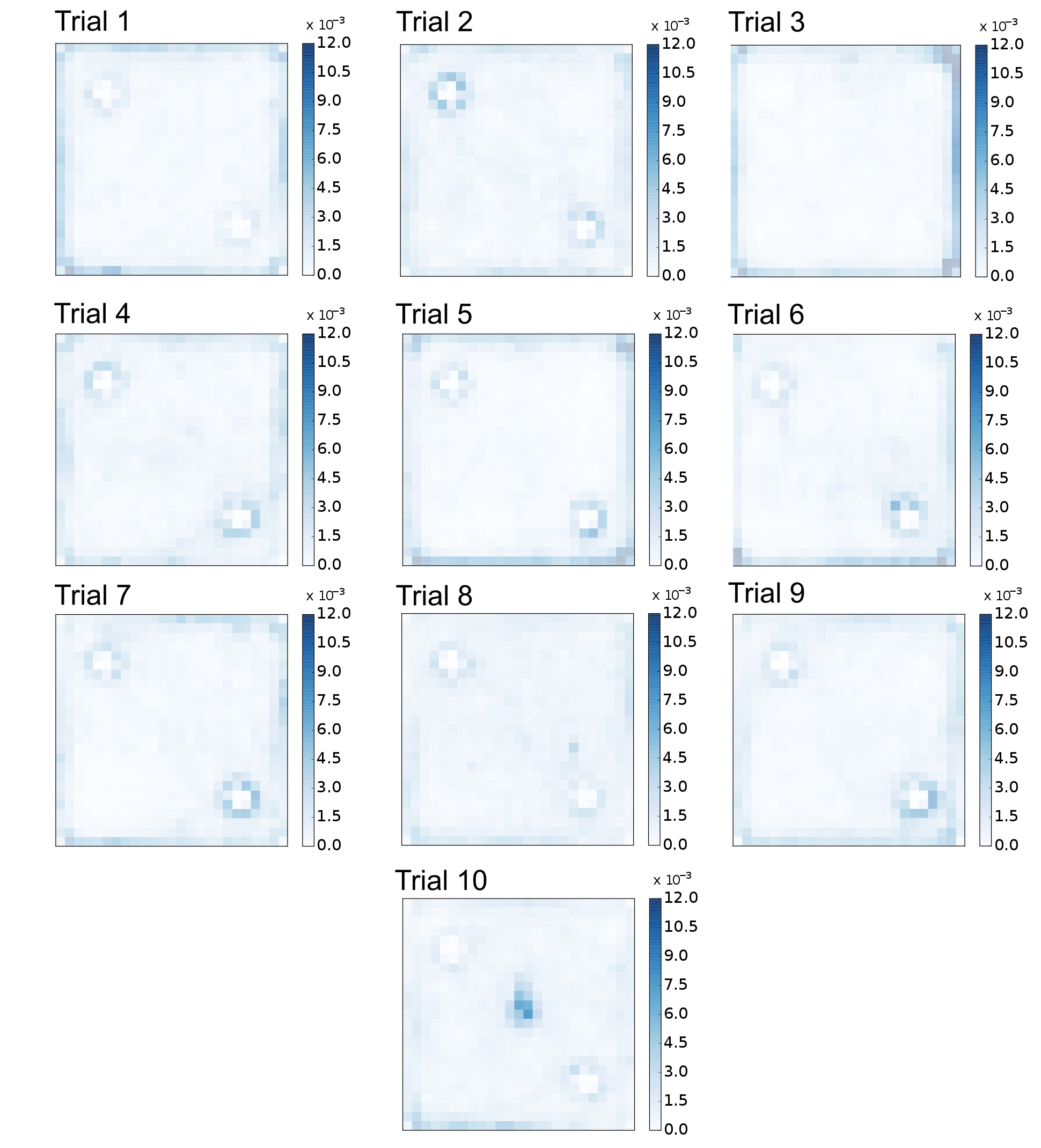}
\caption{\textbf{Probabilities of presence} of 10 trials of 5 TL zebrafish with two cylinders. The probability is calculated on the positions of all zebrafish observed during one hour. The bluer the more fish detected.}
\label{fig:figures3}
\end{figure}

\begin{figure}[ht]
\centering
\includegraphics[width=1\textwidth]{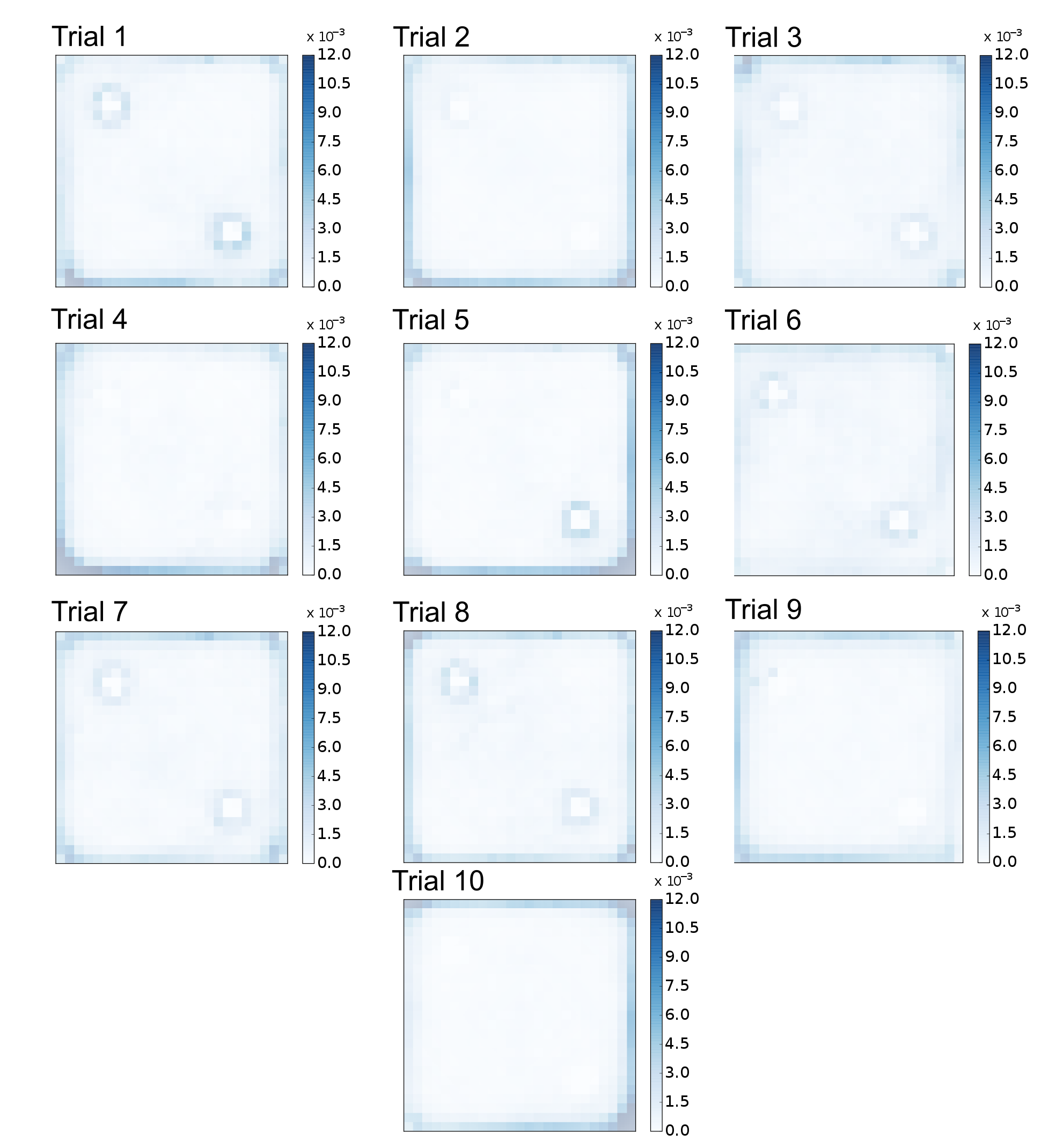}
\caption{\textbf{Probabilities of presence} of 10 trials of 10 TL zebrafish with two cylinders. The probability is calculated on the positions of all zebrafish observed during one hour. The bluer the more fish detected.}
\label{fig:figures4}
\end{figure}

\begin{figure}[ht]
\centering
\includegraphics[width=1\textwidth]{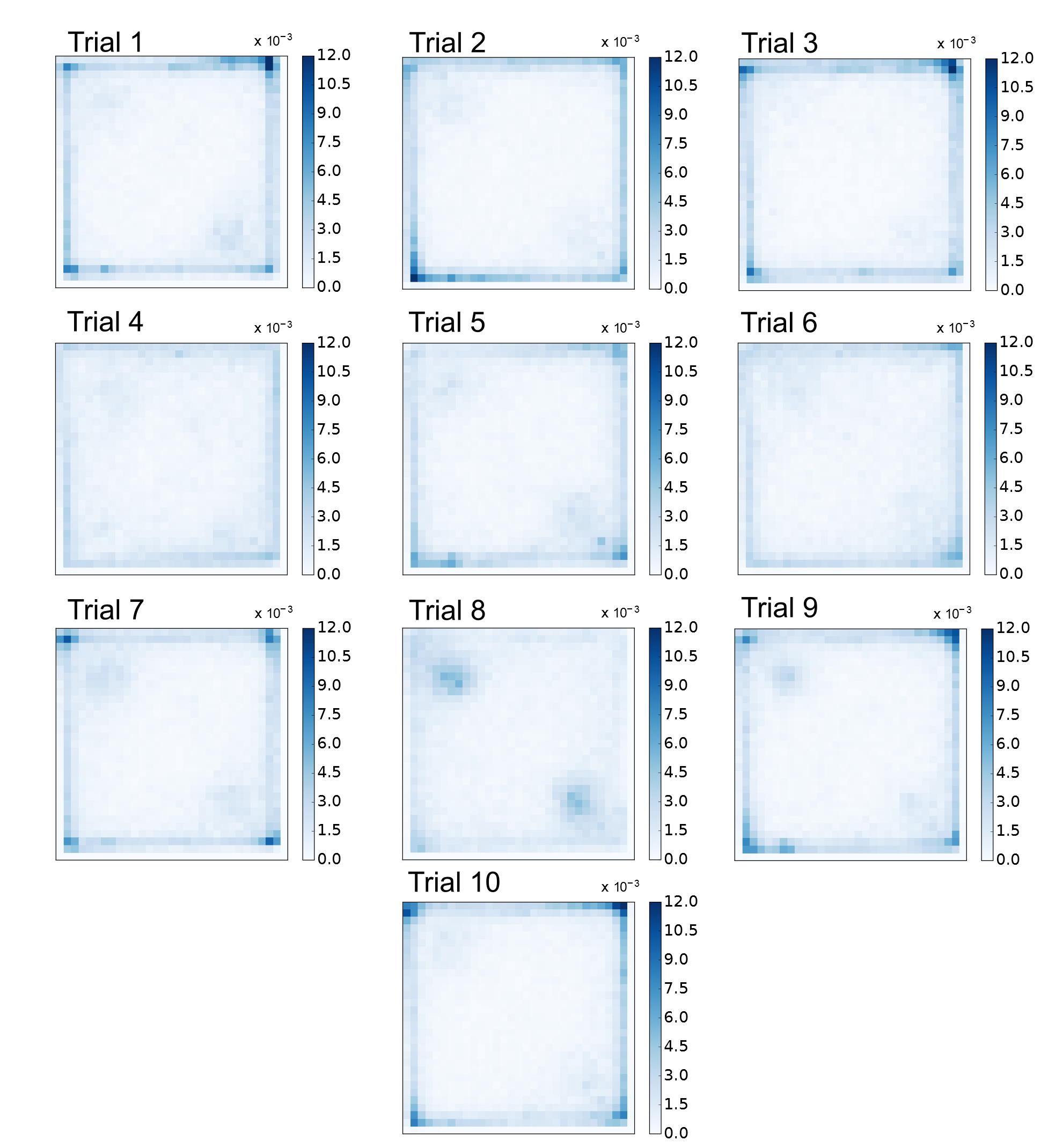}
\caption{\textbf{Probabilities of presence} of 10 trials of 10 TL zebrafish with two disks. The probability is calculated on the positions of all zebrafish observed during one hour. The bluer the more fish detected.}
\label{fig:figures5}
\end{figure}

\begin{figure}[ht]
\centering
\includegraphics[width=1\textwidth]{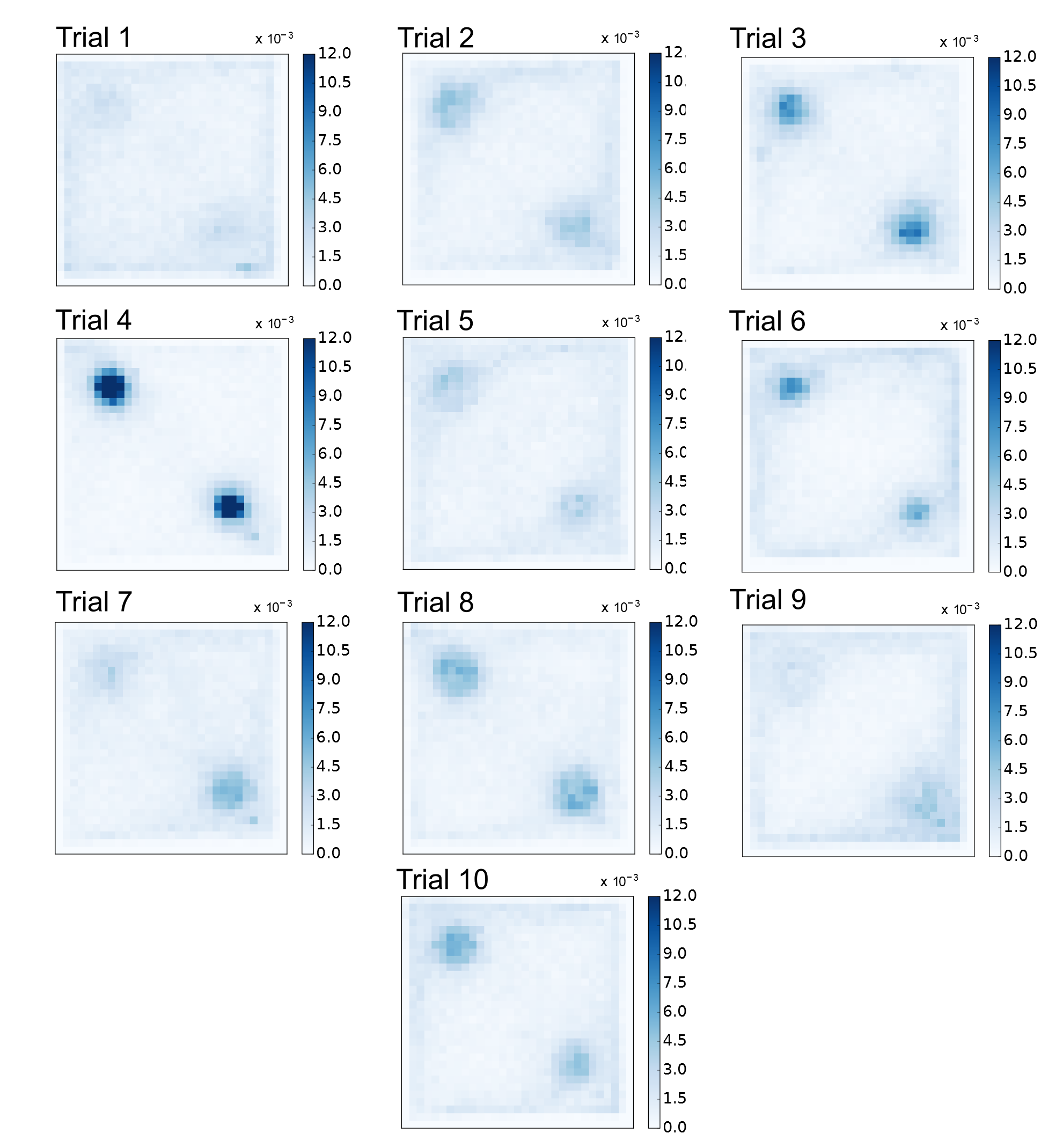}
\caption{\textbf{Probabilities of presence} of 10 trials of 10 AB zebrafish with two disks. The probability is calculated on the positions of all zebrafish observed during one hour. The bluer the more fish detected.}
\label{fig:figures6}
\end{figure}

\begin{figure}[ht]
\centering
\includegraphics[width=0.8\textwidth]{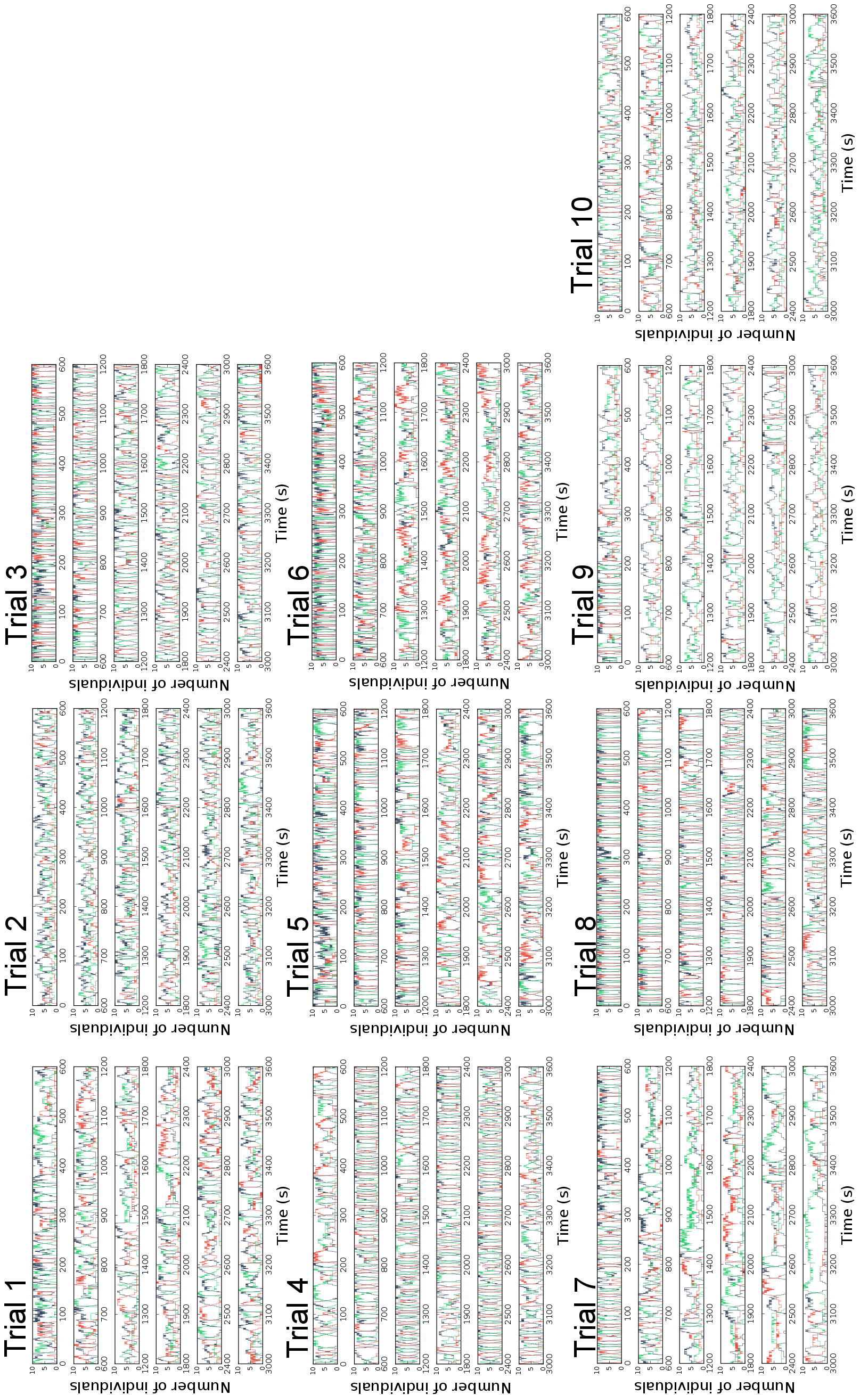}
\caption{\textbf{Landmark occupancy} for 10 trials of 10 individuals of AB zebrafish. For readability, time series are divided in 6 linked subplots. Y-axis reports the number of individuals. Green line represents individuals in the zone of interest (less than 25 cm away from the center of the landmark) of landmark 1, red line the individuals in the zone of interest of landmark 2 and blue line the individuals outside both zones of interest.}
\label{fig:figures7}
\end{figure}

\begin{figure}[ht]
\centering
\includegraphics[width=0.8\textwidth]{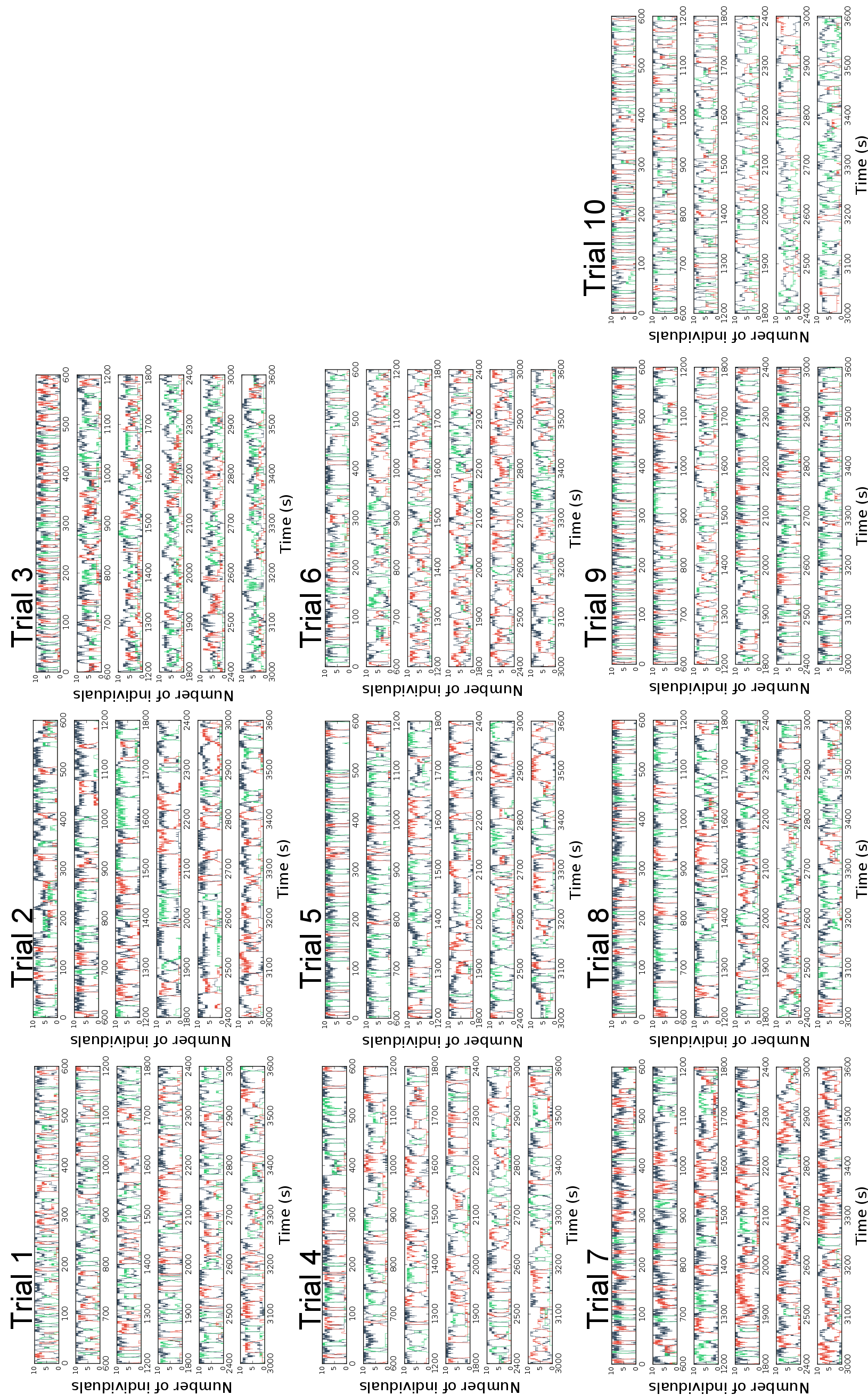}
\caption{\textbf{Landmark occupancy} for 10 trials of 10 individuals of TL zebrafish. For readability, time series are divided in 6 linked subplots. Y-axis reports the number of individuals. Green line represents individuals in the zone of interest (less than 25 cm away from the center of the landmark) of landmark 1, red line the individuals in the zone of interest of landmark 2 and blue line the individuals outside both zones of interest.}
\label{fig:figures8}
\end{figure}

\begin{figure}[h!]
\centering
\includegraphics[width=0.8\textwidth]{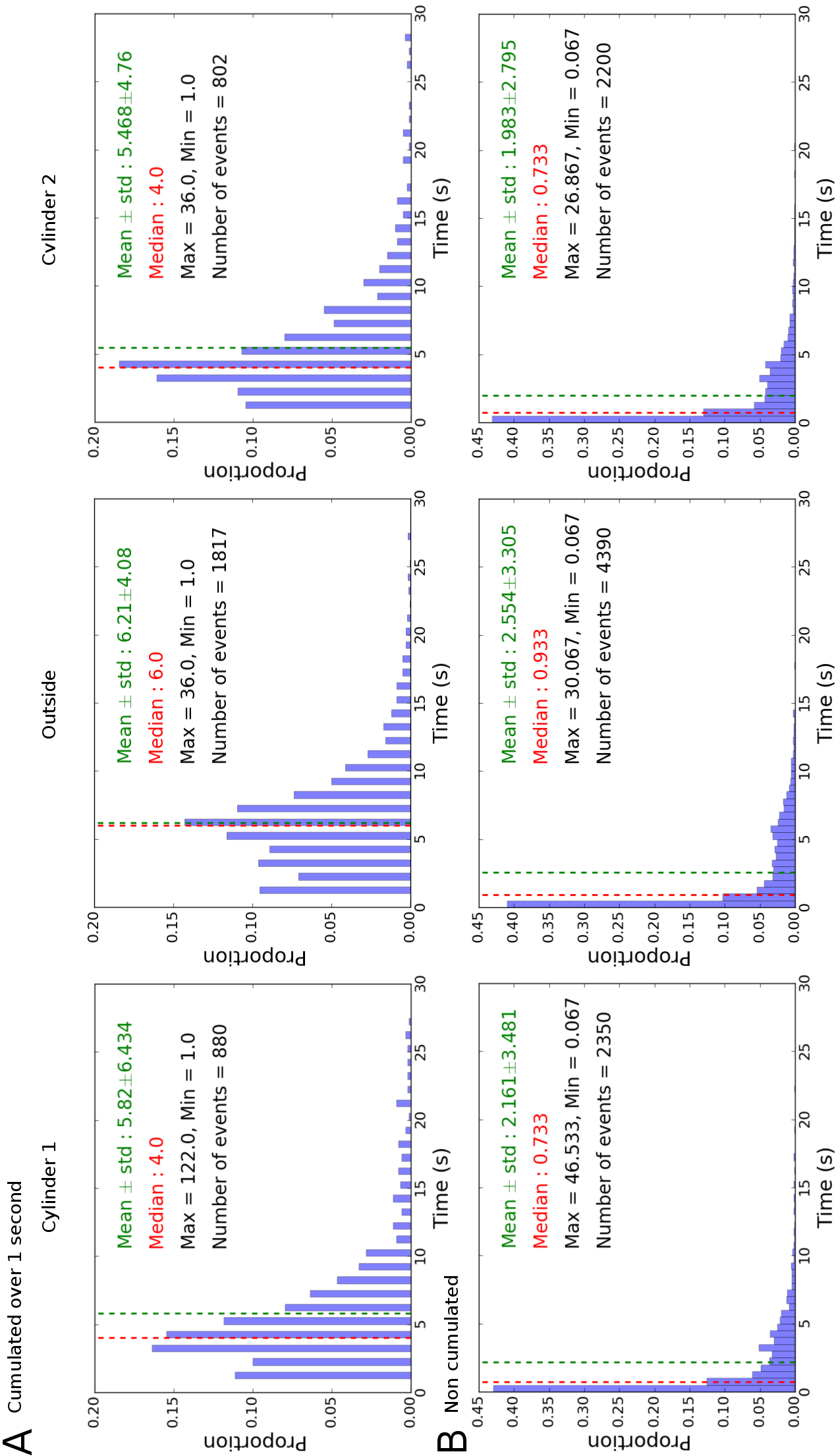}
\caption{\textbf{Comparison of the proportions of the durations of the majority events for AB zebrafish} (A) on cumulated data over 1 second ; (B) on non cumulated data. This figure is related to the Figure 4 of the article. By cumulating the data over 1 second, we decrease strongly the noise. In each area, the median of the presence durations increases and reaches 4 seconds around cylinder 1 and 2 and 6 seconds outside.}
\label{fig:figures9}  
\end{figure}

\begin{figure}[ht]
\centering
\includegraphics[width=0.8\textwidth]{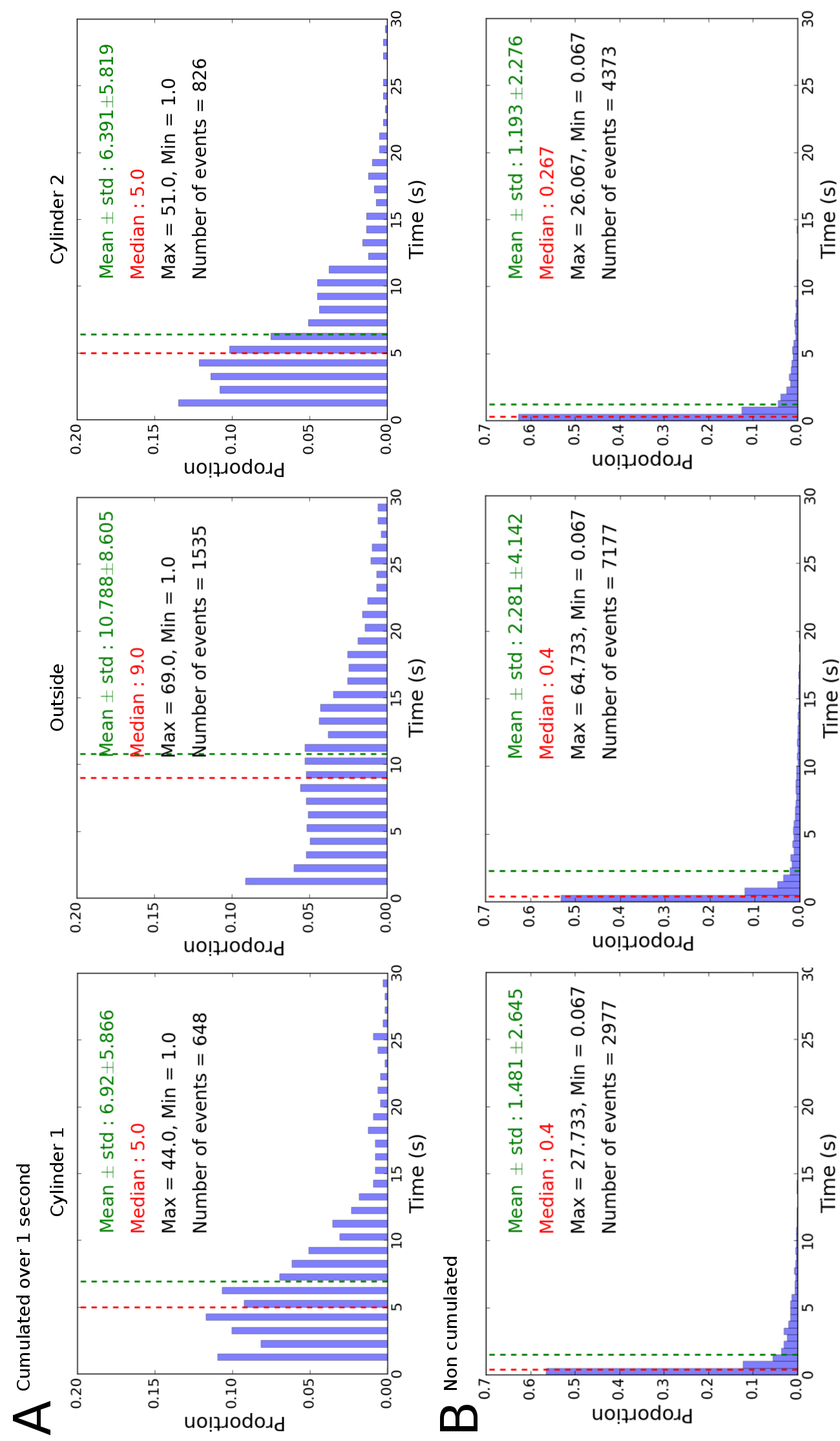}
\caption{\textbf{Comparison of the proportions of the durations of the majority events for TL zebrafish} (A) on cumulated data over 1 second ; (B) on non cumulated data. This figure is related to the Figure 4 of the article. By cumulating the data over 1 second, we decrease strongly the noise. In each area, the median of the presence durations increases and reaches 5 seconds around cylinder 1 and 2 and to 9 seconds outside.}
\label{fig:figures10}  
\end{figure}

\begin{figure}[ht]
\centering
\includegraphics[width=0.8\textwidth]{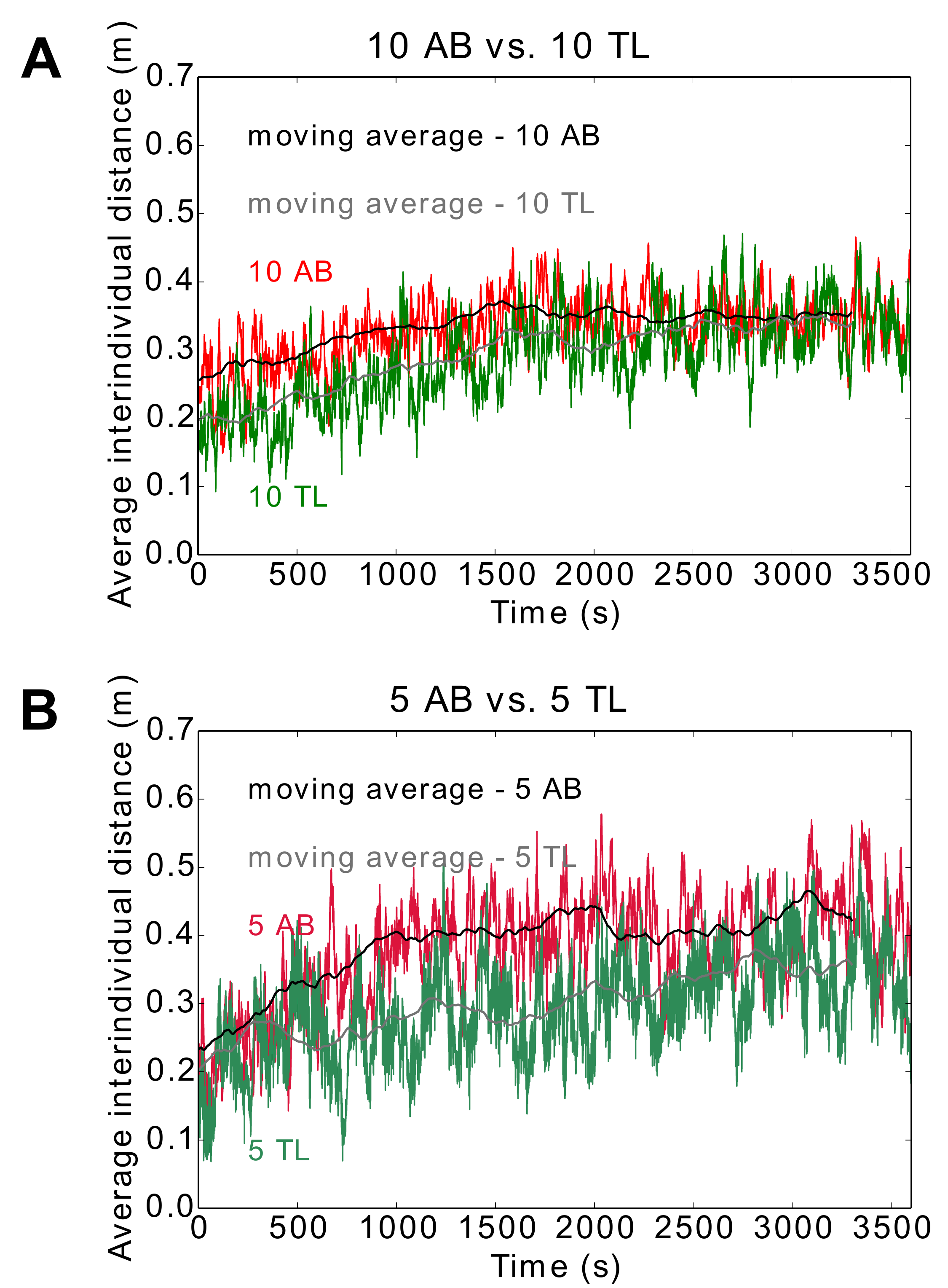}
\caption{\textbf{Time evolution of the average interindividual distance} (A) for 10 AB versus 10 TL fish. The red line is the average of 10 trials with groups of 10 AB zebrafish, the green line is the average of 10 trials with groups of 10 TL zebrafish ; (B) for 5 AB versus 5 TL fish. The red line is the average of 10 trials with groups of 5 AB zebrafish, the green line is the average of 10 trials with groups of 5 TL zebrafish. The average interindividual distance for each condition shows an increase during the first 20 minutes and the reaching of a plateau. It has been calculated as the average of all distances between each couple of fish.}
\label{fig:figures11}  
\end{figure}

\clearpage
\twocolumngrid

\end{document}